\documentclass[pdflatex,sn-mathphys-num]{sn-jnl}

\usepackage{graphicx}%
\usepackage{multirow}%
\usepackage{amsmath,amssymb,amsfonts}%
\usepackage{amsthm}%
\usepackage{mathrsfs}%
\usepackage[title]{appendix}%
\usepackage{xcolor}%
\usepackage{textcomp}%
\usepackage{manyfoot}%
\usepackage{booktabs}%
\usepackage{algorithm}%
\usepackage{algorithmicx}%
\usepackage{algpseudocode}%
\usepackage{listings}%
\usepackage{tabularx}
\usepackage{caption}
\usepackage{makecell}
\usepackage[many]{tcolorbox}
\theoremstyle{thmstyleone}%
\theoremstyle{thmstyletwo}%

\theoremstyle{thmstylethree}%

\newtcolorbox{boxK}{
    sharpish corners, 
    boxrule = 0pt,
    toprule = 0.5pt, 
    enhanced,
    fuzzy shadow = {0pt}{-2pt}{-0.5pt}{0.5pt}{black!35} 
}

\begin{document}

\title{Understanding Developer Pain Points in Federated Learning: Insights from Stack Overflow and GitHub}

\author*{\fnm{Sahand} \sur{Saed}}\email{sahand.saed@usask.ca}
\author{\fnm{Khairul} \sur{Alam}}\email{kha060@usask.ca}
\author{\fnm{Banani} \sur{Roy}}\email{banani.roy@usask.ca}


\affil{\orgdiv{Department of Computer Science}, \orgname{University of Saskatchewan}, \orgaddress{\city{Saskatoon}, \postcode{S7N 5A2}, \state{Saskatchewan}, \country{Canada}}}




\abstract{Federated Learning (FL) enables collaborative model training without centralizing raw data, but building and operating FL systems remains difficult due to distributed execution, rapidly evolving frameworks, and privacy and governance requirements. In this paper, we present an empirical study of FL developer challenges by independently analyzing 495 Stack Overflow posts and 9,116 GitHub issues and pull requests from 92 FL-related projects. Using BERTopic-based topic modeling and difficulty indicators such as unresolved rates and median resolution time, we characterize recurring problem areas and compare how they manifest across the two support platforms, Stack Overflow and GitHub. Our analysis surfaces nine dominant Stack Overflow topics and thirteen GitHub topics, with persistent difficulties concentrated in environment setup and dependency compatibility, API breakages and migration, training instability under non-IID data, evaluation and metric correctness, and the integration of privacy-preserving mechanisms. We also categorize posts by question intent (e.g., “How,” “Why,” “What,” and “other”) to understand the kinds of help developers seek; this intent analysis shows that “How”-type questions dominate, reflecting strong demand for procedural guidance. Several topics such as "TFF Installation \& Environment Compatibility" and "Federated Feature Engineering and SecureBoos Issues" exhibit high unresolved rates and long resolution times, suggesting shortcomings in tooling, documentation, and debugging support. Based on these findings, we provide actionable implications and best practices for FL framework designers, documentation authors, and educators, and we release a replication package to support future research. Although our results are constrained to public discussions and a subset of widely discussed frameworks, the study offers a scalable method for continuously monitoring developer pain points and improving the usability, reliability, and deployability of FL systems.}

\keywords{Federated learning, Developers challenges, GitHub, Stack Overflow}

\maketitle

\section{Introduction}\label{introduction}
Modern machine learning and deep learning have achieved remarkable performance when trained on large, diverse datasets, including in medical image analysis where data scale and label quality strongly influence model generalization \citep{litjens2017survey}. In many real-world deployments, however, data cannot be readily centralized. In healthcare, for example, imaging data are naturally siloed across hospitals and clinics, and sharing can be constrained by privacy, ownership, cross-border transfer requirements, and legal frameworks such as the European Union's General Data Protection Regulation (GDPR) \citep{european2016regulation,hoofnagle2019european}. Beyond governance and compliance, high-quality annotation is often costly and time-intensive in expert-driven domains, further limiting the availability of centralized, labeled datasets \citep{litjens2017survey}. As a result, valuable training data frequently remain distributed across organizations.

Federated Learning (FL) addresses this constraint by enabling multiple parties to collaboratively train a shared model without exchanging raw data. Instead, participants compute local updates and share only model parameters or gradients with an aggregator, reducing the need to move sensitive data off-site. FL was popularized in large-scale mobile settings through the Federated Averaging (FedAvg) paradigm \citep{mcmahan2017communication} and later operationalized in production-grade systems that highlight practical constraints such as device availability, partial participation, and reliability at scale \citep{bonawitz2019towards}. At the same time, FL introduces distinctive challenges relative to centralized learning: statistical heterogeneity from non-IID client data can degrade accuracy and destabilize training \citep{zhao2018federated}, while systems heterogeneity across clients complicates optimization and runtime behavior \citep{li2020federated}. Comprehensive surveys and monographs summarize these algorithmic and system-level challenges and identify open research problems spanning robustness, privacy, efficiency, and evaluation \citep{li2020federated, kairouz2021advances}.

Most deployed FL systems adopt a client--server architecture, where a central coordinator periodically aggregates client updates using FedAvg-like methods \citep{mcmahan2017communication}. Alternative decentralized or peer-to-peer variants remove the central coordinator and rely on neighbor-to-neighbor communication, but they typically introduce additional protocol and operational complexity \citep{lalitha2019peer, hegedHus2021decentralized}. These architectural choices matter because FL must operate under real constraints: intermittent connectivity, stragglers, limited compute and power budgets, and variable client availability, all of which can affect both training dynamics and engineering design \citep{bonawitz2019towards, kairouz2021advances}.

FL is increasingly adopted in privacy-sensitive domains such as healthcare, where multi-institutional collaboration can improve model quality without sharing patient data \citep{Sheller2020_FLinMedicine}. In parallel, FL is gaining traction in IoT and edge settings, motivating a growing ecosystem of open-source frameworks intended to support experimentation and deployment \citep{kholod2020open}. Many real deployments further require privacy-preserving mechanisms beyond keeping data local, such as secure aggregation to hide individual client updates from the server \citep{Bonawitz2017_SecureAgg} and differential privacy techniques to provide formal privacy guarantees \citep{Abadi2016_DPSGD}. Integrating these mechanisms into end-to-end systems introduces additional engineering and debugging complexity \citep{kairouz2021advances}.

Prior research has extensively examined the algorithmic, statistical,
privacy, and system-level challenges of FL \citep{Beutel2020_Flower, He2020_FedML, Reina2021_OpenFL}. These studies have identified
important concerns involving non-IID data, communication efficiency,
client heterogeneity, robustness, privacy, and large-scale coordination.
However, most of this evidence is derived from research publications,
framework descriptions, benchmarks, and controlled experiments. It
provides comparatively limited insight into the day-to-day difficulties
developers encounter when installing, configuring, using, debugging,
deploying, and maintaining FL software.

Empirical software engineering studies have shown that developer
discussions on Stack Overflow and GitHub can reveal practical problems
that may be insufficiently represented in research papers or official
documentation. Prior studies have used these platforms to examine
developer challenges in machine learning, deep learning frameworks,
mobile development, security, and other software domains
\cite{Zhang2018_TFbugs,alam2026analyzing, Humbatova2020_DLFaultTaxonomy,
Alshangiti2019_MLDevChallenges,hamidi2021towards, alam2025empirical,alam2026maintenance}. Nevertheless,
comparable cross-platform evidence remains limited for FL. In
particular, it remains unclear which FL development challenges recur
across public support and repository platforms, what forms of assistance
developers seek, and which topics are associated with unresolved or
time-consuming outcomes.

The scope of this study is limited to developer-reported challenges
visible in publicly available Stack Overflow discussions and GitHub
artifacts from popular open-source FL repositories. Specifically, the
GitHub dataset comprises issues and pull requests from 92 repositories
with at least 300 stars. Accordingly, the findings characterize patterns
observed within prominent public FL development communities rather than
all FL development contexts. They may not represent proprietary systems,
private organizational communication, smaller or less visible
repositories, or discussions occurring through other support channels.

The objective of this study is to empirically characterize the
developer-reported FL challenges visible in public Stack Overflow
discussions and in issues and pull requests from prominent open-source
GitHub repositories. We examine the recurring topics discussed on these platforms, the
intent of the corresponding questions and reports, and the extent to
which different topics exhibit observed resolution difficulty. By
combining evidence from a community question-and-answer platform and
open-source project repositories, the study captures complementary
perspectives on learning, implementation, debugging, deployment, and
software maintenance.

Based on this objective, we investigate the following research
questions:

\begin{itemize}

    \item \textbf{RQ1: What topics are FL developers asking on Stack Overflow and in GitHub projects?}

    This question identifies the recurring technical, operational, and
    support concerns reported by FL developers across the two
    platforms.

    \item \textbf{RQ2: What types of questions are FL developers
    asking?}

    This question examines whether developers primarily seek procedural
    guidance, explanations of unexpected behaviour, clarification of
    concepts or capabilities, or other forms of support.

    \item \textbf{RQ3: To what extent do developers perceive the
    revealed challenges in terms of difficulty?}

    This question examines topic-level resolution outcomes to identify
    areas in which developers may experience limited support or
    time-consuming problem resolution.

\end{itemize}

To answer these questions, we analyzed 495 FL-related Stack Overflow
posts and 9,116 GitHub issues and pull requests collected from 92
public FL-related repositories. We independently applied BERTopic to
the Stack Overflow and GitHub datasets to identify recurring challenge
topics. We then used a validated hybrid rule-based and zero-shot
classification pipeline to categorize the artifacts as How, Why, What,
or Other. Finally, we examined the proportion of artifacts without an
accepted answer or closure and their median resolution time as
complementary indicators of observed resolution difficulty. We also
performed an inferential cross-platform comparison of the
question-intent distributions.

Our analysis identified nine Stack Overflow topics and thirteen GitHub
topics. Stack Overflow discussions primarily reflected user-facing
challenges involving framework implementation, environment
compatibility, data preparation, model-state management, and training
or evaluation. GitHub artifacts more strongly reflected
repository-level concerns involving configuration, reproducibility,
infrastructure, deployment, runtime behaviour, and maintenance. The
question-intent analysis showed that Stack Overflow was more strongly
associated with How-type requests for procedural guidance, whereas
GitHub was more strongly associated with Why-type discussions focused
on explanation and root-cause investigation. The difficulty analysis
further showed that some topics had high unresolved rates, while
others were eventually resolved but required substantial time. These
findings identify areas where improved documentation, compatibility
management, diagnostics, deployment support, and runtime observability
may reduce developer effort.

In summary, our study offers the following key contributions:

\begin{itemize}
    \item \textbf{Empirical Identification and Topic-Based Organization of FL Developer Challenges:}  
    By mining developer discussions from Stack Overflow and GitHub, we empirically uncover real-world challenges encountered during FL implementation and deployment. We then apply BERTopic to cluster posts and issues into coherent challenge topics, yielding a structured taxonomy of the most frequent problem areas that serves as a practical reference for researchers, framework maintainers, and practitioners.
    
    \item \textbf{Question-Type Analysis of Developer Needs:}  
    We analyze the types of questions developers ask on Stack Overflow and find that \emph{``How''-type} questions dominate, indicating a strong need for step-by-step guidance, troubleshooting support, and practical examples. This helps our topic analysis by clarifying what kinds of help developers seek most often.

    \item \textbf{Topic Difficulty Assessment Using Resolution Metrics:}  
    We measure topic difficulty using unresolved rates and median resolution time across Stack Overflow and GitHub. These indicators highlight which topics are most challenging to resolve and provide actionable support for FL developer.
\end{itemize}

\textbf{Paper Organization: }The remainder of this paper is structured as outlined as follows: In Section \ref{background-and-related-work}, we briefly discussed FL, Topic Modeling, and related work. Moving on to Section \ref{study-design}, we detailed our study's methodology. Our findings are presented in Section \ref{case-study-results}, and Section \ref{discussion-implications} illustrates the implications. Section \ref{threats-to-validity} addresses potential threats to the validity of our results. Finally, Section \ref{conclusion} concludes the paper, highlighting directions for future research.

\section{Background and Related Work}\label{background-and-related-work}
FL emerged in response to the growing need to train machine learning models when data cannot be centralized due to privacy, ownership, and regulatory constraints \citep{mcmahan2017communication,kairouz2021advances}. As data generation expanded across domains such as healthcare, mobile devices, finance, and IoT, organizations increasingly faced “data silos,” where valuable datasets remain distributed and sensitive. FL addresses this by enabling collaborative model training without sharing raw data: clients train locally and only share model updates with an aggregation process (e.g., Federated Averaging), which helps reduce direct exposure of private data while still benefiting from multi-party learning \citep{mcmahan2017communication}. Despite major progress in FL algorithms and system designs, building and deploying FL systems remains challenging in practice due to distributed coordination, heterogeneous clients, unstable environments, and rapidly evolving frameworks \citep{kairouz2021advances,bonawitz2019towards}. In this work, we aim to understand FL developers’ real-world challenges by analyzing Stack Overflow posts and GitHub issues and pull requests. In this section, we provide background on FL and summarize prior research on FL systems and developer experience, including work that mines developer discussions to extract recurring pain points and trends.

\subsection{Federated learning prior works and its challenges}
A large body of FL research has focused on algorithms, system architectures, and trustworthy deployment. For example, Kairouz et al.~\cite{kairouz2021advances} provide a broad and influential overview of FL, summarizing core concepts, major challenges, and open problems from the literature, with a strong emphasis on cross-device FL at scale. Beyond such general overviews, several domain-focused surveys review FL in specific application contexts, including healthcare and IoT/edge environments \citep{xu2021federated,nguyen2021federated,nguyen2022federated}. Xu et al.~\cite{xu2021federated} review FL for healthcare informatics and discuss statistical challenges, system challenges, and privacy considerations in medical settings. Their study highlights the need for clearer reporting of practical details such as dataset characteristics, evaluation protocols, and deployment constraints that often determine whether FL solutions can be reproduced and adopted in healthcare.

Other works provide systematic coverage of FL challenges and tooling. Guendouzi et al.~\cite{guendouzi2023systematic} present a systematic review of FL, organizing categories, challenges, aggregation methods, and development tools. Their survey emphasizes recurring technical barriers such as communication overhead, client and data heterogeneity, and privacy requirements, and it summarizes aggregation methods designed to address these issues. Similarly, Lim et al.~\cite{lim2020federated} review FL in mobile edge networks and highlight implementation challenges that arise in large-scale and heterogeneous edge environments, including communication costs, resource allocation, and security concerns.

Several studies also target specialized challenge dimensions. Li et al.~\cite{li2020federated} discuss FL from a broad perspective and outline key challenges that require interdisciplinary solutions spanning distributed optimization, systems, and privacy. Vucinich and Zhu~\cite{vucinich2023current} focus on fairness in FL and summarize motivations, concepts, and open challenges in evaluating and enforcing fairness in FL settings. Shirvani et al.~\cite{shirvani2024federated} review attacks and defenses in FL and argue that security threats (e.g., poisoning and backdoor attacks) remain a major barrier to adoption.

\textbf{Gap and how our work helps.}
Although these prior works provide valuable theoretical and survey-based understanding of FL challenges, they primarily synthesize evidence from published research and do not directly capture the day-to-day difficulties that practitioners face when \emph{installing}, \emph{configuring}, \emph{debugging}, and \emph{maintaining} FL frameworks. In practice, many challenges emerge from environment incompatibilities, API breakages, missing or incorrect evaluation utilities, training instability under non-IID data, and framework-specific implementation details that are often underrepresented in survey papers. Our work complements the existing literature by mining Stack Overflow and GitHub discussions to characterize FL challenges from a \emph{developer-centric} perspective. By extracting topics from posts, issues, and pull requests and analyzing difficulty signals such as unresolved rates and resolution time, our study provides actionable evidence about where developers struggle most and where tooling, documentation, and framework support can be improved.

\subsection{Topic Modeling}

Topic modeling refers to a family of unsupervised learning methods that automatically discover underlying themes (topics) in large collections of unstructured text \cite{jelodar2019latent}. By leveraging patterns of word co-occurrence, these methods can group documents by shared content without manual labeling, which makes them useful for organizing, summarizing, and analyzing large text corpora. As a result, topic modeling has been applied in many areas, including information retrieval \cite{yi2009comparative,wei2007topic}, document clustering \cite{yau2014clustering,xie2013integrating}, recommendation systems \cite{choi2015improving,luostarinen2013using,bergamaschi2015comparing}, summarization \cite{belwal2023extractive}, and machine translation \cite{eidelman2012topic}. It has also been widely used in domains such as software engineering, social media analytics, healthcare, e-commerce, linguistics, and cybersecurity \cite{asuncion2010software,zhai2011constrained,gethers2010using,gokcimen2024topic}.

Classical topic modeling techniques, such as Latent Dirichlet Allocation (LDA) \cite{blei2003latent}, Correlated Topic Models (CTM) \cite{blei2006correlated}, Dynamic Topic Models \cite{blei2006dynamic}, Non-Negative Matrix Factorization (NMF) \cite{lee2000algorithms}, and Biterm Topic Models (BTM) \cite{yan2013biterm}, established the foundation for extracting topics from text. However, these approaches can struggle with short, technical, and context-dependent content, which is common in developer discussions. Recent advances in NLP have enabled embedding-based topic modeling that uses transformer representations to better capture semantics, and BERTopic has emerged as a strong option for producing context-aware and interpretable topics \cite{grootendorst2022bertopic}.

Unlike LDA, which relies on bag-of-words features and typically requires choosing the number of topics in advance, BERTopic uses transformer-based embeddings to represent documents semantically and can derive an appropriate topic structure through clustering. It combines dimensionality reduction (e.g., UMAP) with clustering to form coherent topic groups that remain interpretable even for noisy, large-scale corpora \cite{grootendorst2022bertopic}.

In this study, we used BERTopic to extract and categorize developer challenges from FL-related Stack Overflow posts and GitHub issues and pull requests \cite{grootendorst2022bertopic}. We applied a custom preprocessing pipeline to reduce redundancy and noise, including removing near-duplicate entries and extending standard stopword lists with domain-specific and conversational terms common in developer forums. We generated embeddings using Sentence-Transformers \citep{reimers2019sentence} with the \texttt{all-MiniLM-L6-v2} checkpoint \citep{SentenceTransformers_allMiniLM_L6_v2}, then applied UMAP for dimensionality reduction to support effective clustering \citep{mcinnes2018umap}. To better capture technical expressions, we used a customized \texttt{CountVectorizer} \citep{pedregosa2011scikit} with bi-grams \citep{ScikitLearn_CountVectorizer_Docs}, which improves representation of multi-word terms frequently used in FL discussions.

After processing, we assigned topic labels by extracting representative keywords for each cluster and applied outlier detection to separate posts that did not fit well into any topic. The resulting model supported both quantitative analysis (e.g., topic frequency distributions) and qualitative interpretation through readable topic descriptors. Overall, this workflow enabled a structured analysis of recurring themes and persistent difficulties in FL development communities, while capturing contextual nuances that are often missed by traditional topic modeling methods \cite{grootendorst2022bertopic}.

\subsection{Topic Analysis of Technical Q\&As}

Topic modeling is useful for analyzing technical Q\&A data because it can automatically group large numbers of questions and answers into meaningful categories \cite{chen2023user}. By uncovering hidden themes, it helps organize and index content, making it easier to search and enabling faster discovery of useful knowledge in large and complex datasets \cite{daud2010knowledge}.

On developer platforms such as Stack Overflow and GitHub, where people constantly ask questions and share solutions, topic modeling can cluster related discussions together. This makes it easier to find relevant answers and can reduce repeated questions by highlighting existing discussions \cite{nie2017data}. Topic modeling can also reveal emerging topics, track changes in user interests, and identify common problems or missing knowledge, which helps both users looking for support and tool builders who want to improve their systems \cite{fiscus2002topic}.

Many studies have used topic modeling to analyze Stack Overflow content in general and to study how topics change over time. Others have applied it to specific software domains, including mobile development \cite{rosen2016mobile}, machine learning \cite{alshangiti2019developing}, concurrency \cite{ahmed2018concurrency}, software security \cite{yang2016security}, DevOps \cite{mi2023identifying}, chatbot development \cite{abdellatif2020challenges}, and non-functional requirements \cite{zou2017towards}.

However, to the best of our knowledge, there is still no comprehensive topic-based study that jointly analyzes FL-related discussions across both GitHub and Stack Overflow, even though FL has become an important and fast-growing area. To fill this gap, we use BERTopic because it captures context well and performs strongly on technical text. Using this approach, we extract and analyze key FL topics to provide insights into current trends, frequent developer challenges, and major areas of focus in the FL community.

\subsection{Topic Analysis of GitHub Data}
Topic modeling with GitHub data means analyzing text from issues, pull requests, and comments to identify common themes. This can reveal recurring problems, frequent requests, and emerging trends in software projects. Methods such as LDA and BERTopic help organize this unstructured text into meaningful groups. Researchers use these topics to understand what developers struggle with, what users need, and how project priorities change. Such insights can support better planning, resource allocation, and collaboration, ultimately improving software development practices.

Many studies have applied topic modeling to GitHub data. For example, Li et al.~\citep{li2021understanding} studied challenges in quantum software engineering by analyzing GitHub issues and Stack Overflow posts with LDA. Scoccia et al.~\citep{scoccia2021challenges} used LDA to link GitHub issues with related Stack Overflow questions to identify common web development problems. Other works \citep{alam2025empirical, alam2026analyzing, dhasade2020towards, jokhio2021mining, campbell2015latent, wang2019does} have also used topic modeling on GitHub datasets to uncover developer concerns and trends. Since GitHub issues and pull requests are often short and technical, we used BERTopic \citep{grootendorst2022bertopic}, because its embedding-based approach works well for this type of text.
\newline

To achieve these objectives, the following section outlines the methodology adopted in this study, detailing the data collection process, preprocessing steps, and the implementation of BERTopic for topic extraction and analysis.

\section{Study Design}\label{study-design}
Our study aims to understand what FL developers ask about in practice. To do this, we analyzed developer discussions from two major public sources: Stack Overflow posts and GitHub issues and pull requests. These platforms have been widely used to study real-world developer difficulties in machine learning software, such as the challenges discussed around deep learning frameworks across Stack Overflow and GitHub \cite{han2020deeplearningframeworks}.

\begin{figure}[htbp]
  \includegraphics[width=\textwidth]{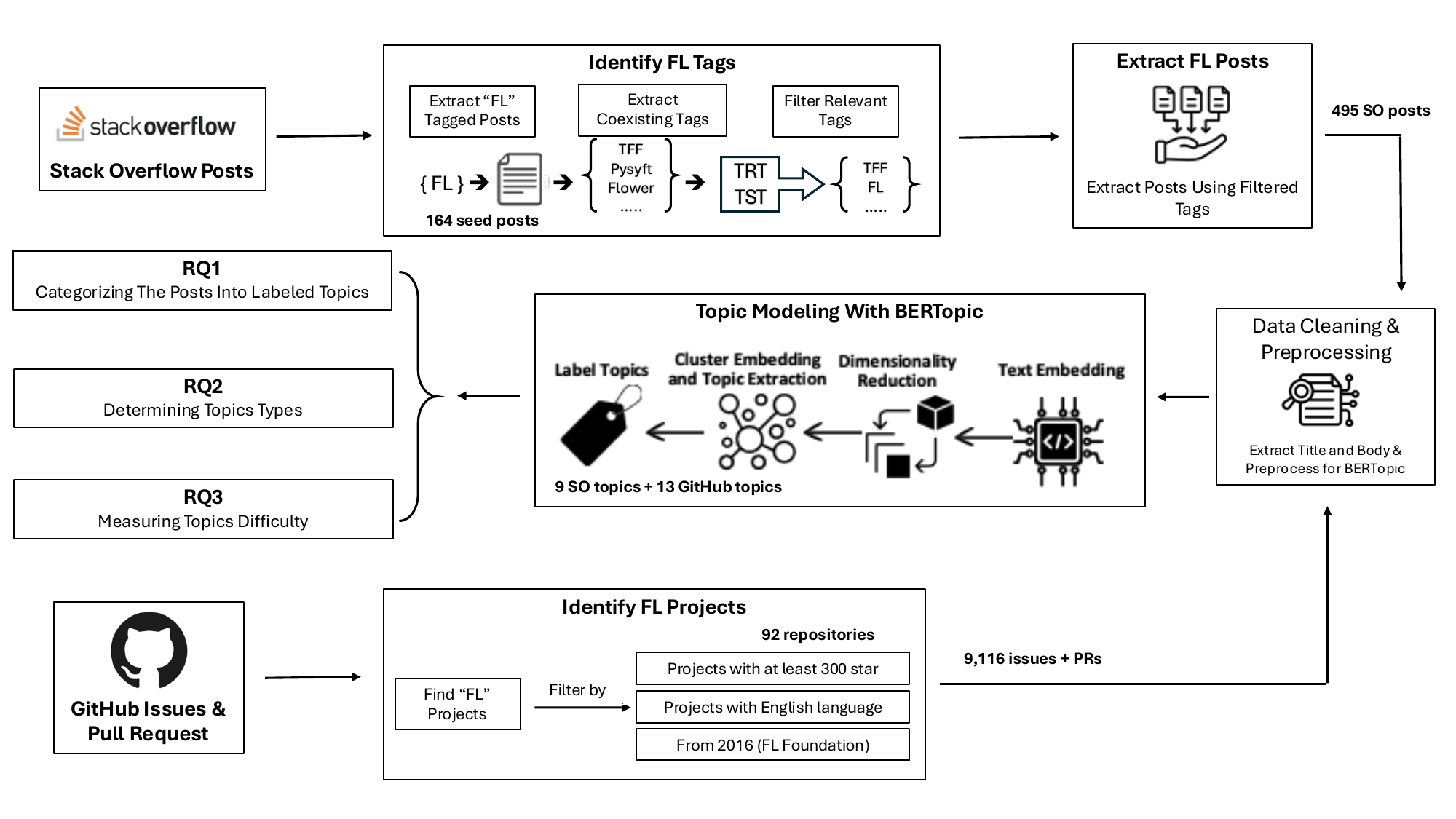}
  \caption{Overview of the methodology of our study}
  \label{fig:studydesign}
  \vspace{-1.0em}
\end{figure}

 In the FL context, developers face additional complexity due to distributed training, heterogeneity, and privacy requirements \cite{li2020federated,kairouz2021advances}. Stack Overflow provides structured Q\&A data and rich metadata (e.g., accepted answers and scores). However, the available tags do not directly yield a stable, analysis-ready taxonomy of FL developer challenges. Therefore, we first identified FL-related posts, grouped them into dominant challenge topics using topic modeling, and then conducted our analyses. For GitHub, we focused on FL-related projects associated with the discussions in our dataset to keep the analysis consistent across both platforms. Figure~\ref{fig:studydesign} summarizes the overall study design, and the following sections describe each step in detail.

\subsection{Data Extraction and Processing of SO data}
In this section, we outline the process of extracting and preparing the Stack Overflow data. Each step is described in detail below.

\subsubsection{How Questions are Asked on Stack Overflow}
Stack Overflow is a Q\&A platform where users can ask questions, post answers, comment, and vote to show which contributions are helpful. To make posts easier to find and organize, users add topic tags to their questions. Each question must have at least one tag and can include up to five tags, which helps categorize the post and improves search results. For instance, a question about building a custom federated dataset in TensorFlow Federated might be tagged \emph{python-3.x}, \emph{tensorflow}, and \emph{tensorflow-federated}.

\subsubsection{Stack Overflow Data Extraction}
\textbf{Step 1: Download \& extract Stack Overflow dump }We downloaded the complete Stack Overflow data dump (last updated on July 26, 2025), which includes user questions, answers, and associated metadata such as view counts and creation dates. This dataset spans the period from August 2008 to July 2026. In total, it comprises 24,212,025 questions and 36,048,359 answer posts.
\newline

\textbf{Step 2: Identifying FL Tag} Stack Overflow hosts a diverse range of software development discussions, covering topics such as Java, Python, cybersecurity, and blockchain. To improve the visibility of their posts and increase the chances of receiving responses, users often assign popular and relevant tags, such as \emph{federated-learning}, \emph{large-language-model}, \emph{chatbot}, and \emph{web}, to their questions \cite{barua2014developers}. To systematically identify the most pertinent tags related to FL, we adopt a methodology inspired by prior studies \cite{abdellatif2020challenges, bagherzadeh2019going, rosen2016mobile}. This involves constructing a curated set of FL-related tags using the following procedure:

First, we retrieve all posts tagged with \emph{federated-learning}, yielding a dataset of 164 posts as of July 26, 2025. To avoid introducing noise at this early stage, we do not include any additional tags, ensuring that this initial dataset remains focused and reliable for identifying related tags. From these posts, we extract all co-occurring tags used alongside \emph{federated-learning}. To systematically expand the set of FL-related tags, we adopt two heuristic metrics from prior research \cite{abdellatif2020challenges, rosen2016mobile, wan2019programmers}. The first is the \emph{Tag Relevance Threshold (TRT)}, which quantifies the strength of association between a candidate tag and the \emph{federated-learning} tag. This is calculated as the ratio of posts containing both the 'federated-learning
' tag and the specific tag to the total number of posts for that tag. Specifically, the TRT is measured using the following equation.
 \[
TRT_{\text{tag}} = \frac{\text{No. of FL posts for the tag}}{\text{Total No. of posts for the tag}}
\] 

We get distinct tags and their occurrence count from the downloaded 164 posts. Then, we count the occurrence of all the tags for the whole dataset using the \emph{tags} table of the SO database. Then we calculate the TRT. For example, the tag 'pysyft' has a TRT of 50\%, indicating that 50\% of posts tagged with 'pysyft' are also tagged with 'federated-learning'. By utilizing the TRT, we can effectively remove irrelevant tags from our tag set. For instance, broad tags such as \texttt{python}, \texttt{pytorch}, and \texttt{docker} typically have low TRT because they frequently appear in non-FL discussions; we therefore exclude them, while FL-specific framework tags (e.g., \texttt{pysyft}) with higher TRT are retained.

However, some tags with a small number of posts (e.g., the 'metadata-repository' tag, which has only 3 posts) may have a high TRT (33.3\%) simply because a single post is related to FL. This can introduce less significant tags into our analysis. To address this, we incorporate a second metric, the \emph{Tag Significance Threshold (TST)}, which measures the prominence of a tag within FL-related posts. This metric is determined by comparing the total number of FL-related posts for a specific tag to the total number of FL-related posts for the most popular tag (the 'FL' tag with 164 posts). The \emph{TST} is calculated as the ratio of FL-related posts for a given tag to the number of posts for the most popular tag. The equation for calculating \emph{TST} is shown below. \[TST_{\text{tag}} = \frac{\text{No. of FL posts for the tag}}{\text{No. of FL posts of the most popular tag}}
\]
For example, the 'tensorflow-federated' tag has a TST of 53.66\%, meaning that the number of posts tagged with both 'tensorflow-federated' and 'federated-learning' is equivalent to 53.66\% of the total number of FL-related posts tagged with 'federated-learning'.

We consider a tag to be significant and relevant to FL-related posts if both its TRT and TST exceed certain thresholds. To determine these thresholds, three individuals with varying levels of FL development experience independently examine tags with different TRT and TST values. For each tag, they review a randomly selected sample of posts to assess when the tag became less relevant or specific to FL. This process aims to identify the most appropriate TRT and TST thresholds, a method employed in several previous studies \cite{abdellatif2020challenges, bagherzadeh2019going, rosen2016mobile}. The goal is to select tags that are relevant to FL while minimizing noise in the dataset.

After evaluating the tags, the three individuals discussed their findings to reach a consensus on the optimal TRT and TST values. They independently assessed the thresholds that yielded the best results and deliberated to agree on a final decision. We find that tags with a TRT higher than 20\% and a TST greater than 10\% yield an appropriate balance between including more FL-related posts such as 'pysyft' (resulting in a more representative dataset) and filtering out posts unrelated to FL (reducing noise). Table \ref{tab:fl_tags} presents the tags included in our tag set along with their corresponding TRT and TST values.

\begin{table}[ht]
\centering
\caption{Top Federated Learning-Related Tag Based on TRT and TST Scores}
\label{tab:fl_tags}
\begin{tabular}{|l|c|c|}
\hline
\textbf{Tag} & \textbf{TRT (\%)} & \textbf{TST (\%)} \\
\hline
federated-learning & 100.00 & 100.00 \\
pysyft & 50.00 & 10.98 \\
metadata-repository & 33.33 & 0.61 \\
tensorflow-federated & 21.95 & 53.66 \\
openehr & 3.45 & 0.61 \\
federated & 2.73 & 1.83 \\
flower & 1.52 & 1.83 \\
ctc & 1.32 & 0.61 \\
johnsnowlabs-spark-nlp & 1.01 & 0.61 \\
ordereddict & 0.85 & 0.61 \\
openfl & 0.73 & 1.22 \\
personalization & 0.62 & 0.61 \\
unet-neural-network & 0.55 & 0.61 \\
loaddata & 0.51 & 0.61 \\
dataloader & 0.47 & 1.22 \\
pytorch-dataloader & 0.42 & 1.22 \\
pytorch-lightning & 0.32 & 1.22 \\
cross-entropy & 0.28 & 0.61 \\
transfer-learning & 0.27 & 1.22 \\
resnet & 0.27 & 1.22 \\
privacy & 0.26 & 1.83 \\
autograd & 0.25 & 0.61 \\
ensemble-learning & 0.23 & 0.61 \\
tensorflow-datasets & 0.19 & 2.44 \\
pytorch & 0.12 & 17.07 \\
\hline
\end{tabular}
\end{table}

Generic tags such as \texttt{python}, \texttt{pytorch}, and
\texttt{tensorflow} were not used independently to retrieve posts
because they contain very large numbers of discussions, most of which
are unrelated to FL. For example, the \texttt{python} tag covers
millions of posts across many software-development domains, while
\texttt{pytorch} and \texttt{tensorflow} also contain substantial
numbers of general machine-learning questions. Retrieving all posts
associated with these tags would therefore introduce considerable noise
and would require manual assessment of a corpus far beyond the scope of
this study. However, posts containing generic tags were not necessarily excluded.
They were included when they also contained an FL-related tag retained
through the co-occurrence, TRT, and TST procedure. Thus, the filtering
strategy uses generic tags as supporting contextual information rather
than treating them as sufficient evidence of FL relevance. Similar
tag-expansion and threshold-based procedures have been used in prior
empirical studies of developer discussions
\cite{rosen2016mobile,abdellatif2020challenges,
bagherzadeh2019going}.

\textbf{Step 3: Extract FL posts.} After obtaining the FL-related tag set, we use these tags to extract the posts that will form the basis of our FL dataset for this study. We gather this corpus by querying all posts on SO that are tagged with any of the tags in our set. This process resulted in a dataset containing 495 unique FL posts along with their respective metadata.
Our dataset can be found in the replication package\citep{saed2026flchallenges}.

\subsection{Stack Overflow Data Preprocessing} 

\textbf{Preprocessing FL posts. } 

We first removed irrelevant content before applying topic modeling. The post Title gives a brief summary of the question, while the Body adds important context and details for identifying the topic. In our analysis, we used both the Title and Body. To reduce noise, we cleaned the Body by removing quotes, HTML tags, links, and code snippets with regular expressions. We also removed common English stopwords, such as 'how', 'can', and 'at', which do not add meaning and may bias results, using the NLTK stopwords list \citep{bird2006nltk}. Finally, we applied lemmatization, which reduces words to their base form, or lemma (for example, 'scale' is the lemma of 'scaling'), while considering their linguistic context. All the data pre-processing steps and processed data are provided in the replication package \citep{saed2026flchallenges}. Below, we provide a concise description of the steps we follow to pre-process the FL-related SO data.
\newline

\subsection{Data Extraction and Processing of GitHub data}
To support our analysis of developer challenges on Stack Overflow, we also looked at issue reports and pull requests from GitHub, a popular platform for software development. These repositories give us a clear view of real problems faced when working with FL. In this part, we explain how we collected and prepared the GitHub data to help identify common topics and challenges.
\newline
\textbf{Step 1: Collecting GitHub Issues and Pull Requests. }
To construct a GitHub dataset suitable for analyzing developer issue discussions in open-source FL projects, we collected data using the GitHub REST API v3. We selected 92 repositories using the following inclusion criteria:

\begin{itemize}
    \item \textbf{Popularity:} at least 300 GitHub stars, indicating an active user community.
    \item \textbf{Time frame and activity:} created or actively maintained since 2016, aligning with the emergence of modern FL frameworks and sustained community adoption.
    \item \textbf{Language:} issues and pull-request discussions predominantly written in English to support consistent text preprocessing and topic modeling.
\end{itemize}

For each selected repository, we retrieved issue and pull-request discussion artifacts, including the title and body text, labels, state (open/closed), creation and update timestamps, author and assignee metadata, comment counts, and the complete comment thread. Data were downloaded via a custom Python pipeline authenticated with a personal access token to respect API rate limits and ensure reliable collection. The raw responses were stored in JSON format and subsequently normalized into CSV tables to facilitate preprocessing and downstream analysis.

In total, our dataset includes over 9116 samples from 92 projects, covering a broad range of FL repositories and development activity.

GitHub issues and pull requests serve related but distinct purposes.
Issues commonly document defects, support requests, feature proposals,
or other project concerns, whereas pull requests propose concrete
changes to code, documentation, configuration, or tests. We retain both
artifact types because together they capture repository-level development
and maintenance activity; however, they are treated as distinct records.
Throughout the manuscript, we use the term \emph{GitHub artifacts} when
referring collectively to issues and pull requests.

As of July 2025, our GitHub dataset comprised a total of 9116 independent issues and pull requests. All scripts used for data collection, PR-issue linkage detection, and filtering are provided in our replication package~\citep{saed2026flchallenges} to ensure transparency and reproducibility.

The 300-star threshold was used to focus on actively used and visible projects with sufficient developer activity. Consequently, the GitHub findings primarily
characterize challenges reported in popular open-source FL frameworks
and may not represent smaller, newer, specialized, or proprietary FL
projects.
\newline
\textbf{Step 2: Preprocessing GitHub Data. }To prepare the GitHub issues and pull requests for topic modeling, we applied a series of preprocessing steps to clean and normalize the textual data. First, we concatenated the title and body of each issue and pull request to form a single document, ensuring sufficient context for topic extraction. We then removed code snippets enclosed in backticks using regular expressions, as such blocks often contain syntax elements irrelevant to semantic modeling. Next, we eliminated URLs, which typically reference external documentation or repositories, and cleaned out any remaining HTML tags using regular expressions. To further reduce noise, we removed special characters and punctuation. We also applied stopword removal using the NLTK stopword list\cite{hardeniya2016natural}. Finally, we performed lemmatization using \emph{spaCy’s en\_core\_web\_sm} model \cite{vasiliev2020natural} to convert words into their base forms, thus improving topic coherence and reducing lexical variation. 

\subsection{Identify Federated Learning's topics. } \label{identify-topics}
BERTopic starts with transforming input documents into numerical representations. While there are several methods to accomplish this, \emph{SentenceTransformers}\cite{mezinijumping} stands out as a state-of-the-art technique for generating sentence and text embeddings. Renowned for its ability to capture semantic similarities between documents, it is among the most popular choices for this task. There are many pre-trained models available for \emph{SentenceTransformers}, all hosted on the Huggingface Model Hub \cite{huggingface-models-2025}. The \emph{all-*} models were trained on an extensive dataset of over one billion training pairs, making them suitable for general-purpose applications. In addition, there are \emph{Multi-QA} models trained on 215 million question-answer pairs from diverse sources, including StackExchange, Yahoo Answers, and search queries from Google and Bing. Notable models in this category include \emph{multi-qa-MiniLM-L6-dot-v1, multi-qa-distilbert-dot-v1, and multi-qa-mpnet-base-dot-v1}. The \emph{multi-qa-mpnet-base-dot-v1} model excels in semantic search performance, whereas the \emph{multi-qa-MiniLM-L6-dot-v1} model is optimized for speed. We chose to use the \emph{multi-qa-MiniLM-L6-dot-v1} model for our embedding needs because it is trained on question-answer data, which closely aligns with the nature of our dataset.

In BERTopic, the \emph{nr\_topics} parameter allows control over the number of topics by merging similar ones after their initial creation. However, best practices recommend using a clustering model to determine the topic structure more naturally \cite{li2021understanding}. To enable accurate and coherent clustering, we first reduced the dimensionality of the document embeddings, as high-dimensional data can impair clustering performance. Among dimensionality reduction techniques, \emph{UMAP} \cite{mcinnes2018umap} is widely recognized for its ability to preserve local structure when projecting data to lower dimensions. Therefore, we employed UMAP for this step. We then applied \emph{HDBSCAN} \citep{mcinnes2017hdbscan}, a density-based clustering algorithm that complements UMAP well due to its sensitivity to local structure. Unlike traditional clustering methods, HDBSCAN can identify noise and does not force every data point into a cluster, making it especially effective for short, noisy texts such as SO posts and GitHub issues

For BERTopic modeling on both datasets, we carefully tuned UMAP parameters to balance topic granularity and semantic coherence. We experimented with \emph{n\_neighbors} values between 15 and 50 to ensure that semantically similar documents were grouped together while preserving a meaningful global topic structure. The \emph{n\_components} parameter was adjusted between 3 and 20 to retain essential semantic information from the high-dimensional sentence embeddings while enabling efficient clustering. We used \emph{cosine distance} as the metric, as it aligns well with sentence-transformer embeddings and effectively captures angular similarity. For \emph{HDBSCAN}, we tested \emph{min\_cluster\_size} values from 50 to 300 in intervals of 10. We employed \emph{Euclidean distance} for clustering, as it performs reliably in UMAP-reduced space, which is optimized for Euclidean geometry. Additionally, we used the \emph{eom} method for \emph{cluster\_selection\_method} because it identifies stable, well-separated clusters by emphasizing dense core regions. We improved the default representation of topics using the \emph{Countvectorizer} \footnote{\href{https://scikit-learn.org/stable/modules/generated/sklearn.feature\_extraction.text.CountVectorizer.html}{https://scikit-learn.org/stable/modules/generated/sklearn.feature\_extraction.text.CountVectorizer.html}}. It helps to ignore infrequent words and increase the n-gram range. Following previous works \citep{li2021understanding, abdellatif2020challenges}, we used the unigram and bigram models in Countvectorizer. The optimal topic coherence score of 0.64 was achieved for the GitHub dataset with \emph{n\_neighbors} = 20, \emph{n\_components} = 4, and \emph{min\_cluster\_size} = 130, while the SO dataset yielded the optimal score of 0.63 with \emph{n\_neighbors} = 30, \emph{n\_components} = 3, and \emph{min\_cluster\_size} = 210. This configuration enabled BERTopic to generate high-quality, interpretable topics while maintaining scalability across both dataset.

Using the above-mentioned parameters, we obtained nine topics for SO data and 13 topics from the GitHub data. BERTopic conventionally assigns the label \emph{-1} to documents that
cannot be confidently associated with the identified clusters and are
therefore treated as outliers. In our analysis, extensive preprocessing
removed noisy, duplicated, and irrelevant textual content before topic
modelling. We subsequently inspected the documents assigned to
\emph{-1} and found that they represented a coherent and recurring theme
rather than a miscellaneous set of outliers. We therefore retained this
group as a substantive topic and relabelled all topics sequentially for
presentation. Accordingly, the original \emph{-1} cluster is reported as
the first topic in our results. The preprocessing and topic-generation
scripts for the Stack Overflow and GitHub datasets are available in the
replication package \citep{saed2026flchallenges}. The resulting topics
are presented in the following section.

It is important to clarify that BERTopic serves as the initial organizing component of our methodology rather than the sole analytical lens. We use the model to group semantically related Stack Overflow posts and GitHub artifacts into recurring FL development topics. The automatically generated topic representations are then interpreted and validated through manual examination of representative artifacts and topic keywords by the authors.

The resulting topics provide the units for several complementary analyses. We measure topic prevalence to identify frequently discussed challenges and use platform-specific engagement measures to examine the attention received by Stack Overflow topics. We classify questions into How, Why, What, and Other categories to determine the forms of assistance developers seek. We further assess topic difficulty using the proportion of posts without accepted answers or unresolved GitHub artifacts and their median resolution time. Finally, we examine topic distributions over time and compare the findings across Stack Overflow and GitHub to distinguish user-facing learning and troubleshooting concerns from repository-level implementation, maintenance, and deployment challenges. Thus, topic modelling identifies the underlying thematic structure, while the subsequent manual, descriptive, difficulty, temporal, and cross-platform analyses support the interpretation of developer pain points from multiple perspectives.

\subsection{Topic Visualization}
To complement the tabular presentation of the GitHub topics, we
visualized their semantic relationships using principal component
analysis (PCA)~\cite{jolliffe2016principal}. For each topic, we calculated a
centroid by averaging the sentence embeddings of all GitHub issues and
pull requests assigned to that topic. We then applied $\ell_2$
normalization to the topic centroids and projected them onto the first
two principal components. In the resulting visualization, each bubble
represents a topic, and its size is proportional to the number of
assigned artifacts. Spatial proximity provides an approximate indication
of similarity between topic-centroid representations.

\section{Results}\label{case-study-results}
In this section, we presented the analysis of the FL-related data we obtained from SO and GitHub and the topics to answer our research questions.
\newline
\subsection{RQ1: What topics are FL developers asking on Stack Overflow and in GitHub projects?} 

\textbf{Motivation: } Understanding the topics raised in issues and pull requests is essential for identifying the challenges and needs in FL framework development and usage. Because FL frameworks are often community-driven and actively evolving, their issue trackers capture recurring technical difficulties, usability concerns, feature requests, and collaborative development practices. Importantly, the development and deployment of FL systems differ substantially from those of traditional centralized software systems. FL frameworks must coordinate distributed clients, manage heterogeneous data partitions, support diverse learning paradigms (e.g., horizontal, vertical, and decentralized FL), and operate under strict privacy and communication constraints. They must also function reliably across a wide range of execution environments, including standalone simulations, multi-party deployments, edge devices, and cloud platforms.

These requirements introduce domain-specific challenges—such as client heterogeneity, non-IID data distributions, secure aggregation, differential privacy integration, system orchestration, and reproducibility across distributed settings that are far less prevalent in general-purpose software engineering. As a result, the challenges faced by developers and users of FL repositories differ fundamentally from those encountered in conventional software projects. The motivation behind this research question is therefore to uncover the dominant challenges, unmet needs, and recurring concerns reflected in FL-related issues and pull requests, providing a structured understanding of where FL frameworks succeed and where they struggle in practice.
\newline
\textbf{Approach: } We used BERTopic to identify the main topics
developers discuss about FL on Stack Overflow and GitHub. The analysis
produced nine topics from the Stack Overflow data and 13 topics from the
GitHub data, together with the key co-occurring words representing each
topic. We then assigned clear and meaningful labels to the identified
topics. Following methods used in prior studies~\cite{li2021understanding,
bagherzadeh2019going}, the first author with over two years of experience in FL proposed an initial label for each
topic based on its keywords and a manual examination of
randomly sampled artifacts, including posts, issues, and pull requests.
Because the number of artifacts varied substantially across topics, we
did not use a fixed sample size. For larger topics, we examined between
20 and 30 artifacts, while for smaller topics, we reviewed at least ten
artifacts. The topic keywords, representative artifacts, and sampled
discussions were considered together to ensure that each proposed label
captured the dominant theme of the corresponding topic. The second author, who has over five years of similar research experience,
reviewed the proposed labels and supporting artifacts and discussed
potential refinements with the first author. When the two authors
encountered ambiguity or could not reach consensus, the topic was
referred to the third author, who has more than ten years of empirical research
expertise, for additional review and resolution. After incorporating the
feedback from both co-authors, the first author finalized the labels for
all topics. This iterative topic-labeling and validation process was one
of the most time-intensive stages of the study. We also used topic
frequencies to identify the most prevalent FL topics discussed by
developers on each platform. To do this, we used two complementary
popularity measures from earlier research~\cite{abdellatif2020challenges,
bagherzadeh2019going, ahmed2018concurrency, mezinijumping,
bajaj2014mining}:

\begin{enumerate}
    \item \textbf{The average number of views (avg. views) }of the post received from both registered and unregistered users serves as a valuable metric. A high view count implies significant interest and popularity among FL developers. This metric effectively measures community engagement by indicating how frequently a post is viewed.
    \item \textbf{The average score (avg. scores)} of the posts received from the users serves as another important metric for measuring popularity. SO allows its members to up vote posts they find interesting and valuable and down vote posts to maintain high content standards. These votes are aggregated into a score, serving as a metric of the post's perceived community value.
\end{enumerate}

\renewcommand{\arraystretch}{2}
\begin{table*}[htbp]
    \centering
    \caption{FL topics, keywords, and their popularity for Stack Overflow data.}
    \label{tab:swssotopics}
    \begin{tabularx}{\textwidth}{r|>{\raggedright\arraybackslash}p{3.5cm}|>{\raggedright\arraybackslash}p{5cm}|r|r|r}
        \toprule
        SL. & Topic & Keywords & \# Posts & AvgView & AvgSc \\
        \midrule
        1 & TFF Implementation \& Client & client, model, function, federated, datum, running. & 167 & 494.53 & 0.98 \\
        2 & TFF API \& Compatibility Errors & error, code, function, input, package. & 76 & 686.17 & 1.08 \\
        3 & FL Convergence \& Evaluation Mismatch & accuracy, training, round, result, loss. & 55 & 394.45 & 1.01 \\
        4 & TFF Model State \& Aggregation & weight, state, send, save, update. & 53 & 264.98 & 0.98 \\
        5 & TFF Installation \& Environment Compatibility & install, version, try, import, dependency. & 48 & 770.37 & 0.60 \\
        6 & Data Partitioning and Loading for FL &  dataset, split, create, set, get, select. & 40 & 606.48 & 1.48 \\
        7 & PySyft Setup \& API Breakages & pysyft, hook, attribute, virtualworker, configuration. & 21 & 1860.57 & 1.28 \\
        8 & DP in FL (Clipping \& Noise) & privacy, clip, noise, value, gradient, label, party. & 20 & 224.70 & 0.80 \\
        9 & FL Frameworks – Scope \& Deployment Reality & support, network, multiple, simulate, real, machine. & 15 & 468.53 & 1.13 \\
        \bottomrule
    \end{tabularx}
\end{table*}
\renewcommand{\arraystretch}{2}

\textbf{Results Obtained Using SO: }Following the process described in sub-section \ref{identify-topics}, we identified nine topics for SO data. Table \ref{tab:swssotopics} presents the obtained nine topic titles along with their associated keywords. It also shows the number of posts related to each topic and the popularity of these topics based on our metrics: views and scores received by developers on SO. As shown in Table \ref{tab:swssotopics}, developers inquire about various topics in FL development, with the number of posts varying across topics. Below, we discussed these topics in greater detail.

\textbf{1. TFF Implementation \& Client:}
\noindent
This topic describes the practical challenges developers face when trying to run FL correctly with TensorFlow Federated (TFF) and TensorFlow/Keras. Many problems come from fitting real models and real data pipelines into TFF's strict computation rules, where inputs must match precise type and shape requirements (e.g., \texttt{input\_spec} and \texttt{TensorSpec}). Code that works in standard TensorFlow often fails once it is wrapped in \texttt{@tff.tf\_computation} or \texttt{@tf.function} and executed through TFF.

\par\medskip
A common source of difficulty is building and preprocessing federated datasets, including data conversion, batching, handling \texttt{SparseTensor} inputs, non-IID partitioning, and client-local data access under distributed execution. Developers also struggle to adapt existing Keras models into TFF, especially when models have non-standard interfaces (e.g., multi-input models), rely on checkpoints, or require custom prediction and weight-handling logic. In addition, TFF introduces runtime and scalability challenges such as executor configuration, serialization limits for large models, unexpected CPU/GPU behavior, and out-of-memory failures when scaling the number of clients. Overall, this topic highlights the gap between FL ideas and the practical constraints of TFF, where success depends on carefully aligning model and data interfaces with TFF's execution and placement requirements.

\textbf{2. TFF API \& Compatibility Errors: }
This topic summarizes common TensorFlow Federated (TFF) ``plumbing'' failures that prevent developers from running FL code reliably. The main causes are (1) version incompatibilities across the TFF--TensorFlow--Python stack, which lead to missing or renamed APIs and import-time errors (e.g., \texttt{AttributeError}, \texttt{TypeError}, \texttt{ValueError}), and (2) TFF's strict type and shape requirements, where datasets and model inputs must match exact \texttt{TensorSpec} structures and federated type signatures. Even when installation succeeds, small mismatches in dataset structure, batch dimensions, Keras model interfaces, or execution context (eager vs.\ graph, context reuse) can cause opaque failures. Overall, the topic highlights that successful FL experimentation in TFF often depends on carefully aligning library versions and precisely matching TFF's input specifications before any algorithmic work can begin.

As a representative illustration, one common failure pattern occurs when a dataset produced from tabular data is passed into a federated training loop and TFF raises a \texttt{TypeError} because the provided tensors do not match the expected \texttt{TensorSpec} (e.g., a missing or inconsistent batch dimension such as \texttt{float32[15]} versus \texttt{float32[32]}). This example highlights a recurring issue in this topic: even if a data pipeline works in standard TensorFlow, TFF requires dataset element structure and shapes to match the federated type signature exactly, otherwise execution fails with low-level specification errors.

\textbf{3. FL Convergence \& Evaluation Mismatch: }
This topic captures a recurring \emph{``it runs, but it does not learn''} failure mode in TFF. Developers report that training executes for many rounds and metrics are logged normally, yet the global model shows little to no improvement: accuracy stays near chance, improves only on some clients, or collapses to predicting a single class. Discussions commonly point to two root sources. First, training and evaluation pipelines may be misaligned---for example, differences in preprocessing, label encoding, \texttt{input\_spec} definitions, batching, or how evaluation data are constructed can make the reported training progress incomparable to validation behavior. Second, federated optimization can behave poorly under realistic conditions: non-IID client data, participation of too few clients per round, too many local epochs, or poorly tuned learning rates can cause client updates to cancel out or push the server model toward a degenerate solution. Developers also note that the problem is more pronounced for deeper architectures, especially those using Batch Normalization, and for larger models where hyperparameters that work in centralized training do not transfer reliably to federated training.

For instance, in the post \textit{``model performance improve FL training''}, the developer observes that federated training with FedAvg completes normally and the loss decreases across rounds, suggesting apparent progress. However, after each aggregation round they convert the server state back into a Keras model and evaluate it on a held-out validation set, only to find that the model repeatedly predicts a single class and the validation performance does not improve. They emphasize that the same XceptionNet-based architecture performs well under centralized training on the same task, making the federated outcome surprising. The troubleshooting discussion focuses on whether the restored global weights are actually applied correctly when reconstructing the Keras model from server state, whether evaluation uses the exact same preprocessing and label mapping as training, and whether the \texttt{input\_spec} used to build the TFF model matches the dataset element structure and batching used during evaluation. It also raises federated-specific factors---such as heterogeneous client distributions, partial participation, an overly large client learning rate, or too many local epochs---that can lead to unstable or canceling updates and a degenerate global model despite decreasing training loss.

\textbf{4. TFF Model State \& Aggregation: }
This topic focuses on TFF ''wiring layer'' challenges, where the main difficulty is controlling how model weights, updates, and state move between \texttt{CLIENTS} and \texttt{SERVER}. Developers frequently encounter friction when attempting tasks that are straightforward in standard TensorFlow---such as saving models (\texttt{.h5}), initializing from pretrained checkpoints, extracting client-side models, or inspecting and modifying update tensors---because TFF represents values as placement-aware objects that exist inside federated computations rather than as ordinary Python objects. As a result, discussions often revolve around how to access and persist the server state (including optimizer state), how to resume training correctly, and how to transfer weights between Keras and TFF in both directions. The topic also includes requests for more advanced behaviors, such as communication-efficient updates (e.g., sparsification or compression), custom aggregation rules beyond FedAvg, and stateful methods (e.g., per-client state or SCAFFOLD-style tracking). Overall, it highlights that deeper customization in TFF requires working within its strongly typed, placement-aware programming model rather than relying on standard Python control flow and Keras utilities.

For instance, in the post \textit{``save train TensorFlow Federated model .h5 model ?''}, a developer trains a model using Federated Averaging and then attempts to save the trained result as a standard Keras \texttt{.h5} file, expecting the same workflow as centralized training. They also ask how to access both the aggregated global model after training and the individual client models in order to inspect how each client changed during local updates. The ensuing discussion surfaces a key issue in this topic: the primary artifact returned by \texttt{tff.learning.algorithms.build\_weighted\_fed\_avg}-style training loops is a server-side state object at \texttt{SERVER} placement, while client models and updates exist at \texttt{CLIENTS} placement inside federated computations. Consequently, directly calling \texttt{model.save()} on the federated output fails or produces confusing errors because the object being returned is not a Keras model but a placement-aware state container. The resolution emphasizes that exporting a \texttt{.h5} model requires explicitly mapping weights from the server state into a concrete Keras model instance, and that extracting per-client models similarly requires a placement-aware computation that materializes client-specific weights, rather than relying on ordinary Python-side saving and inspection calls.

\textbf{5. TFF Installation \& Environment Compatibility: }
This topic summarizes the setup and environment problems developers face before they can run any FL experiment with TensorFlow Federated (TFF). The main issue is strict version compatibility: TFF releases are tied to narrow ranges of Python and TensorFlow, which often conflicts with existing environments or preinstalled stacks (e.g., Google Colab). Common symptoms include pip resolver warnings, import-time errors (often Keras/typing-related), forced restarts, and missing wheels on local machines (``no matching distribution found''), especially on Apple Silicon and Windows. Even when installation succeeds, runtime failures can occur due to notebook quirks (e.g., kernel freezes, \texttt{asyncio} conflicts) or code that assumes source builds rather than pip packages. Overall, this topic highlights that version pinning, packaging gaps, and uneven platform support can make ``install $\rightarrow$ import'' a major barrier before any FL work begins.

One illustrative report appears in the post \textit{``\_module keras.api.v2.keras.experimental attribute peepholelstmcell''}. The developer explains that \texttt{pip install} completes successfully in Google Colab, but importing TFF fails immediately at \texttt{import tensorflow\_federated as tff} with a Keras attribute error referencing \texttt{peepholelstmcell}. The failure happens before any training code is executed, and the traceback points to a missing or unexpected symbol under \texttt{keras.api.v2.keras.experimental}. The discussion connects this behavior to a compatibility mismatch between the TensorFlow/Keras stack preinstalled in the notebook runtime and the dependency versions expected by the chosen TFF release. Consequently, the workflow breaks at the install/import stage despite a seemingly successful installation, demonstrating how version pinning and packaging constraints can prevent developers from reaching even a minimal federated experiment.

\textbf{6. Data Partitioning and Loading for FL: }
This topic summarizes TensorFlow Federated (TFF) ``data plumbing'' challenges that arise when developers use custom datasets instead of built-in \texttt{ClientData}. TFF requires each client to provide a \texttt{tf.data.Dataset} with a consistent element structure and \texttt{TensorSpec}, but real data often arrives as per-client files (e.g., CSVs) or as non-TFF formats (NumPy arrays, Python iterables, Keras tuples). Developers therefore struggle to convert siloed data into TFF-compatible client datasets, perform correct per-client train/test splits, and ensure consistent preprocessing and batching across clients. Additional friction includes handling unbalanced client sizes, saving/reloading federated datasets, and avoiding workflows that break \texttt{tf.data} semantics (e.g., converting to NumPy or trying to append samples during training). Overall, reliable TFF training depends on a well-defined \texttt{dataset\_fn}, consistent dataset specifications across clients, and preprocessing that stays compatible with \texttt{tf.data} and TFF's federated type system.

A representative example appears in the post \textit{``load fashion MNIST dataset Tensorflow Fedarated ?''}. The developer explains that EMNIST runs successfully because it is already packaged as \texttt{tff.simulation.ClientData}, meaning each client can directly produce a \texttt{tf.data.Dataset} with the expected element structure. In contrast, Fashion-MNIST loaded through Keras is returned as NumPy arrays (or Keras-style tuples), which do not satisfy TFF's requirement for per-client \texttt{tf.data.Dataset} objects with consistent \texttt{TensorSpec}s. The discussion therefore focuses on how to convert the Keras-provided arrays into a TFF-compatible federated dataset: defining client partitions, wrapping each client's subset via \texttt{tf.data.Dataset.from\_tensor\_slices}, applying consistent preprocessing and batching across clients, and ensuring that the resulting element structure matches the \texttt{input\_spec} used to build the TFF model. The post explicitly contrasts the out-of-the-box convenience of \texttt{ClientData} with the manual conversion and specification work required for custom datasets, directly reflecting the core challenge captured by this topic.

\textbf{7. PySyft Setup \& API Breakages: }
This topic summarizes ``PySyft cannot even start'' problems, where developers are blocked before running any FL logic due to ecosystem and version instability. Common issues include failed installations and dependency conflicts (especially strict PyTorch/\texttt{torchvision} pinning that no longer matches Colab/Kaggle defaults), import-time \texttt{ModuleNotFoundError} caused by packaging or internal layout changes (e.g., missing \texttt{syft} or \texttt{syft\_proto} modules), and API churn where older tutorials no longer match newer releases (e.g., missing \texttt{TorchHook}, \texttt{VirtualWorker}, or \texttt{FederatedDataLoader}). Some failures also occur at runtime in websocket-based setups due to connection lifecycle issues. Overall, the topic highlights that PySyft experimentation often requires careful version pinning and alignment between tutorial code, dependencies, and the execution environment.

One representative report appears in the post \textit{``Syft (PySyft): ModuleNotFoundError syft\_proto.messaging.v1.protocol\_pb2''}. The developer attempts to run a minimal PySyft snippet in Google Colab, but the workflow fails immediately at import time with a \texttt{ModuleNotFoundError} referencing an internal \texttt{syft\_proto} path. The traceback indicates that the expected \texttt{syft\_proto} package (or a specific generated protobuf module under it) is not present in the installed environment, even though the user followed standard installation steps. The ensuing discussion points to packaging and internal layout changes across PySyft versions, where modules referenced by older code or older releases are renamed, relocated, or removed, and where dependency drift (e.g., mismatched PyTorch/\texttt{torchvision} versions in Colab) further increases the chance of a broken install. As a result, the developer is blocked before any federated training logic can run, illustrating how import-time failures in the surrounding ecosystem can make ``getting to the first working example'' a central pain point for PySyft users.

\textbf{8. Differential Privacy in FL (Clipping \& Noise): }
This topic summarizes the difficulties developers face when enabling differential privacy (DP) in FL pipelines, especially in TensorFlow Federated or libraries such as Opacus. The main challenges are not simple configuration errors, but the strong interactions between clipping/noise, optimization dynamics, and evaluation. Developers report large utility drops, poor generalization, and even divergence when DP is enabled, showing how sensitive DP-FL is to clip norms, noise multipliers, learning rates, batch sizes, local epochs, and client participation. They also struggle with conceptual questions about where clipping happens (client vs.\ server, per-example vs.\ per-batch, global vs.\ layer-wise) and how to perform privacy accounting in round-based FL (mapping DP-SGD parameters to $(\varepsilon,\delta)$ under partial participation). Additional friction comes from dynamic privacy schedules and confusion about local versus central DP. Overall, the topic highlights that DP is not a plug-and-play feature in FL; achieving both privacy and utility requires careful parameter tuning, correct accounting, and consistent evaluation.

A representative example is shown in the post \textit{``apply Differential Privacy TensorFlow Federated''}. The developer reports that the same model trains normally when DP is disabled, but after enabling DP aggregation---introducing gradient clipping and a \texttt{noise\_multiplier}---the training becomes unstable: loss increases each round and validation performance degrades rather than improves. They experiment with multiple clipping thresholds and noise settings, yet the divergence persists, prompting questions about which component is responsible for the instability and where clipping is applied in the TFF DP pipeline. The discussion highlights the coupled nature of DP and optimization in federated settings: adding noise changes the effective update scale, interacts with the client optimizer and learning rate, and can be amplified by non-IID data and limited client participation per round. As a result, the workflow demonstrates a central pain point captured by this topic---enabling DP can fundamentally change training dynamics, and resolving the failure typically requires coordinated tuning of privacy parameters and optimization hyperparameters, along with careful, consistent evaluation and privacy accounting.

\textbf{9. FL Frameworks – Scope \& Deployment Reality: }
This topic summarizes a common scope and expectation gap around TensorFlow Federated (TFF). After running FedAvg-style tutorials in notebooks, developers often ask whether TFF is mainly a simulation/research tool or can be used for real multi-device deployments. Many questions focus on what TFF provides (federated computations and learning logic) versus what must be built externally (device management, networking, long-running clients, and handling unreliable connectivity). Developers also ask about extensibility beyond FedAvg (e.g., Federated SGD, FedProx, hierarchical FL), and about compatibility with non-neural models (e.g., SVMs, tree models, reinforcement learning) that may not fit TFF's differentiable-update assumptions. Additional confusion concerns what counts as ``federated'' in simulations (client sharding, horizontal vs.\ vertical FL, communication realism) and how to write code that works both in standard TensorFlow and inside TFF's federated execution model. Overall, the topic highlights uncertainty about TFF's intended use, extensibility limits, and the infrastructure needed to move from tutorials to real deployments.

A representative example appears in the post \textit{``Tensorflow Federated -- Learning simulating federate learn one machine?''}. After completing common TFF tutorials, the developer asks whether the code they ran is merely simulating FL on a single machine or whether TFF can be used to train across multiple machines with real, networked clients. The question leads to a discussion that separates TFF's core functionality---defining federated computations and learning processes---from the surrounding infrastructure required for deployment, such as orchestrating client devices, maintaining long-running client processes, handling communication, and coping with unreliable connectivity and partial participation. The thread also raises practical concerns about how far the tutorial-style FedAvg workflow generalizes to other algorithms and settings, and whether the same code structure can be reused outside a notebook environment. In this way, the post directly reflects the central issue of this topic: developers are uncertain about TFF's intended scope and what additional components are necessary to move from a simulated tutorial to an operational federated system.

\textbf{Results Obtained Using GitHub: }We identified 13 key topics developers encountered when working with GitHub data associated with FL following the process described in sub-section \ref{identify-topics}. Our analysis of GitHub issues and pull requests revealed that these challenges often stem from the inherent complexity and interdisciplinary nature of these systems.

\begin{table*}[htbp]
\footnotesize
    \centering
    \caption{Topics derived from FL on GitHub}
    \label{tab:sws_topics_github}
      \begin{tabularx}{\linewidth}{r|>{\raggedright\arraybackslash}p{7cm}|>{\raggedright\arraybackslash}p{6.5cm}}
        \toprule
        SL. & Topic & Keywords \\
        \midrule
        1 & FL Deployment \& Operational Failures & line, error, client, code, module, test. \\
        2 & Federated Data Partitioning \& Pipelines & dataset, vertically, example, partition datum. \\
        3 & FL Debugging, Configuration, \& Operations & psi, intersection, altimage, pulsar. \\
        4 & FL Benchmark Reproducibility Issues & result, run, algorithm, apfl, help, code. \\
        5 & Federated Wiki Docs, Navigation \& Collaboration & page, wiki, fork, add, make, sfw. \\
        6 & Training instability \& evaluation failures & epoch, line, acc, train, average, fedavg, false. \\
        7 & FATE Flow Operations, CLI, \& Data-Management & job, fateflow, component, cli, standalone. \\
        8 & Federated Feature Engineering \& SecureBoost Issues & bin, relate, onehot, hetero feature, secureboost, optimize, inference, host.\\
        9 & FL Capability Gaps Across Settings & vertical, node, xgboost, implement, y0. \\
        10 & KubeFATE Deployment \& Operations Troubleshooting & kubefatee, mysql, vluster, info, run, mode use, compose. \\
        11 & Docker pull/build/run issues & pull, build, docker run, admin, unknown, desription \\
        12 & GPU device \& memory misconfiguration & fedmlclient, nvflare, gpu memory, default value, memory use. \\
        13 & Runtime \& RPC Failures & line, object, cur, recent file, true, traceback, job statursunne. \\
        \bottomrule
    \end{tabularx}
\end{table*}

\begin{figure*}[htbp]
    \centering
    \includegraphics[
        width=\textwidth
    ]{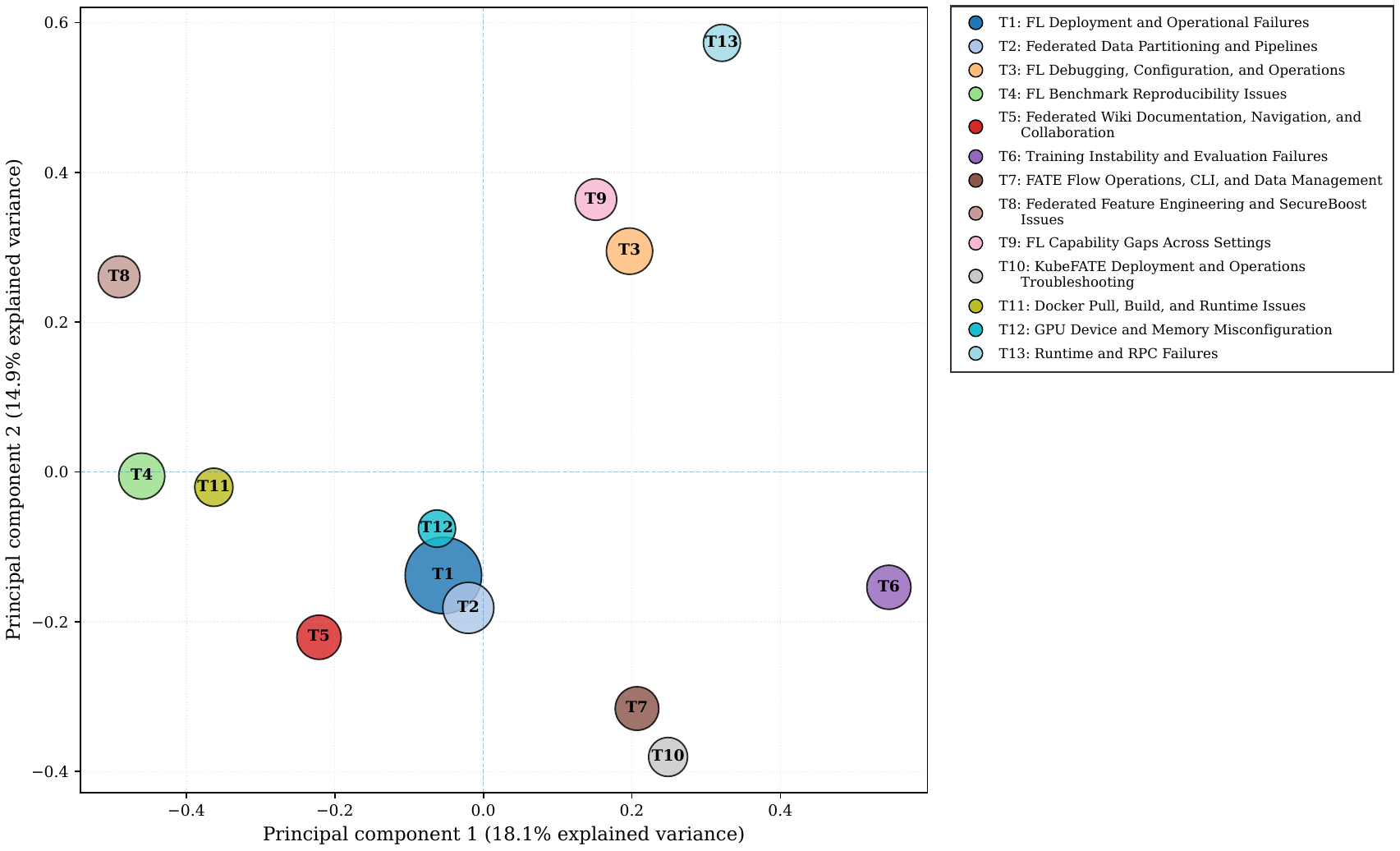}
    \caption{PCA projection of the GitHub topic centroids. Each bubble
    represents a topic, and its size is proportional to the number of
    assigned GitHub issues and pull requests. Spatial proximity provides
    an approximate indication of similarity between the normalized
    topic-centroid embeddings. The first two principal components
    explain 33.0\% of the total variance.}
    \label{fig:github_topic_pca}
\end{figure*}

Figure~\ref{fig:github_topic_pca} complements Table~\ref{tab:sws_topics_github}
by providing a two-dimensional overview of the relationships among the
GitHub topics. The first and second principal components explain 18.1\%
and 14.9\% of the variance, respectively, accounting for 33.0\% in
combination. The projection indicates several local groupings. T1
(\textit{FL Deployment and Operational Failures}), T2
(\textit{Federated Data Partitioning and Pipelines}), and T12
(\textit{GPU Device and Memory Misconfiguration}) occupy a closely
positioned region. T3 (\textit{FL Debugging, Configuration, and
Operations}) is positioned near T9 (\textit{FL Capability Gaps Across
Settings}), while T7 (\textit{FATE Flow Operations, CLI, and Data
Management}) is located near T10 (\textit{KubeFATE Deployment and
Operations Troubleshooting}). By contrast, T6, T8, and T13 appear more
separated from the other topic centroids in the projected space.

These positions provide an exploratory indication of similarities and
distinctions among the topic representations. Because the first two
components retain only part of the variance in the original embedding
space, the visualization is interpreted together with the representative
keywords, descriptions, and examples reported in
Table~\ref{tab:sws_topics_github}.

\textbf{1. FL Deployment and Operational Failures:}
This topic summarizes the practical engineering problems developers face when running distributed FL systems, focusing on making deployments work reliably rather than on the learning algorithm. Common issues include installation and dependency conflicts across Python/TensorFlow/PyTorch (e.g., \texttt{numpy} mismatches, \texttt{ModuleNotFoundError}, \texttt{mpi4py} build failures) and platform-specific failures on Windows, macOS, Linux, Jetson, and Raspberry Pi. Developers also struggle with configuration mistakes (e.g., \texttt{FileNotFoundError}, mis-set arguments or environment variables, misconfigured components such as \texttt{KrumFusionHandler} or DP modules). Networking and communication errors (gRPC/MPI/MQTT, port and protocol mismatches, cloud/edge restrictions) frequently halt training, and runtime failures such as segmentation faults, memory errors, and hangs are hard to debug due to limited visibility across distributed components. Overall, the topic highlights the gap between FL algorithms and dependable end-to-end execution in real environments.

One illustrative discussion is titled \textit{``running or upgrading FedML on an NVIDIA Jetson device''}. The developer explains that they are blocked not by the FL training logic, but by the Jetson software stack's strict version constraints and hardware-specific dependencies. When attempting to install or upgrade FedML, they encounter a requirement to tightly align JetPack with CUDA/cuDNN versions, and to match those with compatible PyTorch/\texttt{torchvision} builds and numerical libraries such as NumPy. The thread emphasizes that even small mismatches in this dependency chain can prevent the environment from building or importing correctly, leading to failures before distributed training can start. The discussion also reflects the compounding nature of platform constraints on specialized devices: upgrading one component (e.g., PyTorch) can force cascading upgrades or downgrades across the stack, making reproducible deployment difficult. In this way, the post highlights a central theme of this topic---on edge hardware such as Jetson, deployment reliability and dependency alignment can dominate the practical effort required to run FL systems.

\textbf{2. Federated Data Partitioning \& Pipelines: }
This topic summarizes the data-integration challenges in FL, where developers struggle to obtain datasets, convert them into framework-specific formats, and partition them into client/party views that match the intended setting (horizontal FL, non-IID FL, vertical FL, or split learning). Common problems include broken download links, missing files or checksum mismatches, slow loaders, unclear dataset requirements, and confusion about where data-related configuration is defined. Developers also report uncertainty about preprocessing steps (often inconsistent with papers/README files) and difficulty creating correct IID/non-IID splits or controlling the number of clients when loaders implicitly determine it. Vertical FL and split learning add further complexity because data must be aligned across parties, sometimes requiring identifiers or private set intersection. These challenges are amplified by inefficient data pipelines and documentation gaps around schemas, normalization, and end-to-end dataset onboarding. Overall, the topic highlights that reliable and reproducible FL depends heavily on clear dataset definitions, robust preprocessing, explicit partitioning logic, and well-aligned documentation.

A typical example is the post \textit{``Which methods are used to preprocess the data set?''}. Rather than asking about the learning algorithm, the developer is blocked by uncertainty about the required preprocessing pipeline: which cleaning, normalization, feature-construction, or filtering steps must be applied, and in what order, to reproduce the reference workflow. The question also asks how the dataset should be transformed so that it can be partitioned correctly into the intended party views, including preparation steps needed for settings such as SplitNN or vertical FL where alignment across parties is required. The thread reflects a recurring pattern in this topic: even when an FL method is conceptually clear, ambiguous or under-specified preprocessing requirements make it difficult to prepare data in the expected schema and to create the intended client/party partitions. As a result, experiments become hard to reproduce and execute reliably because the data pipeline---not the FL algorithm---becomes the primary source of friction.

\textbf{3. FL Debugging, Configuration, \& Operations: }
This topic summarizes the ``make it run and make it match the paper'' stage of FL development, where developers face practical problems beyond the algorithm itself. Common issues include unexpected framework behavior, configuration settings that do not take effect (e.g., YAML parameters being ignored or overridden), and difficulties running pipelines in realistic environments such as multi-party deployments, Kubernetes, and cross-version setups. Developers often see a mismatch between what they configure (e.g., client counts) and what the framework actually executes, which forces source-code and log inspection and reduces trust in experimental control. Reproducibility concerns are frequent, including uncertainty about whether implementations match papers, how randomness and initialization are handled, and how to interpret hyperparameters and evaluation procedures. Deployment complexity adds further friction through networking, platform-specific compatibility (ARM/macOS/Linux), and integration of privacy/security components such as PSI. Overall, the topic highlights that dependable FL systems require reliable configuration semantics, stable framework behavior, deployable infrastructure, and clear alignment between documentation, code, and experimental expectations.

One concrete illustration is the post \textit{``fednlpThe clientnumintotal which I set in fedmlconfig.yaml doesn't work''}. The developer sets \texttt{clientnumintotal} and related parameters such as \texttt{clientnumperround} in a YAML configuration file, expecting the framework to instantiate a specific number of clients and follow the corresponding sampling behavior during training. When they run the pipeline, however, the logs and runtime behavior indicate that a different effective client count is being used. The discussion explores why the configured values are not reflected in execution, raising possibilities such as the YAML file being ignored, values being overridden by command-line arguments or defaults, or the framework implicitly deriving the client count from dataset partitioning logic. To diagnose the discrepancy, the developer is pushed toward inspecting source code paths and configuration-loading order, as well as tracing how dataset loaders construct client partitions. The post therefore illustrates a core pattern in this topic: configuration-semantic mismatch undermines experimental control and makes it difficult to trust that the intended FL setup---and by extension, the intended comparison to a paper---is actually what the system runs.

\textbf{4. FL Benchmark Reproducibility Issues: }
This topic summarizes experiment-execution problems in FL, where developers encounter NaNs, crashes, missing parameters, broken or incomplete examples, slow or stuck runs, incorrect evaluation, and results that do not match papers. The core concern is whether an FL method can run end-to-end, produce meaningful metrics, and be reproduced reliably. Common failure modes include numerical instability (NaN/Inf values, division-by-zero, segmentation faults), brittle scripts that fail due to missing arguments or incomplete experiment drivers (e.g., \texttt{experiments.py}), and reproducibility gaps caused by missing details about hyperparameters, preprocessing, seeds, dataset splits, and evaluation settings. Developers also report evaluation bugs (e.g., incorrect confusion matrices, wrong evaluation order, missing inference outputs) and performance bottlenecks (e.g., inefficient dataloaders, stuck Kubernetes pods, NCCL issues, silent OOM). Overall, the topic highlights that ``code runs'' is not enough; stability, correct evaluation, and reproducibility are required for trustworthy FL results.

One illustrative case is the post \textit{``Miss many arguments in most of the experiments.''}. The developer attempts to run the provided experimental scripts as intended for end-to-end evaluation, but execution fails immediately because multiple entry points call functions without supplying required arguments. Rather than encountering an algorithmic limitation, the user is blocked by incomplete experiment drivers: the scripts cannot even reach the stage where training finishes and metrics are computed. The discussion focuses on identifying which parameters are missing, where defaults are assumed but not implemented, and how the experiment runner should be wired so that configuration values are passed consistently into the training and evaluation pipeline. Because the code fails before producing outputs, the developer cannot validate metrics, compare with reported results, or reproduce claims from the paper. The post therefore captures a central message of this topic: execution correctness and complete experiment scaffolding are prerequisites for meaningful evaluation and reproducibility in FL.

\textbf{5. Federated Wiki Docs, Navigation \& Collaboration: } 
This topic summarizes documentation- and wiki-related challenges where the main difficulty is understanding and using the system through pages, wikis, and guides rather than fixing FL algorithms or runtime errors. Users struggle with outdated or missing documentation, broken links and 404s, and unclear quick-start instructions. Many questions involve wiki mechanics (navigation, revision history, forking/merging, collaboration workflows) and technical details such as \texttt{localStorage} and local-only edits. Usability problems include confusing URLs, redirects, and difficulty locating content, alongside requests for clearer onboarding, stronger API docs, fixed documentation builds (e.g., Sphinx/ReadTheDocs), consistent docstrings, and updated end-to-end examples. Deployment and offline workflows add further friction when users run local instances and face plugin/CORS issues. Overall, the topic shows that successful adoption depends on reliable documentation infrastructure, clear navigation, and an understandable collaboration model, not just correct documentation text.

A concrete illustration is the issue titled \textit{``Improve documentation for compute package development.''}. Rather than reporting a bug in an FL algorithm, the author explains that they are blocked by missing or unclear guidance on how the system is intended to be used and extended. The discussion calls for documentation that makes the development workflow explicit, including how contributors should structure and publish changes, how revision information and metadata should be interpreted, and what collaborative actions such as forking and merging mean within the wiki's collaboration semantics. The thread also highlights navigation and discoverability problems---users cannot reliably locate the relevant instructions due to incomplete pages and scattered guidance---which forces them to infer expected behavior from trial-and-error and fragmented wiki content. In this way, the post aligns with the core theme of this topic: documentation infrastructure and collaboration clarity can be the primary barriers to adoption even when the underlying software is functional.

\textbf{6. Training instability \& evaluation failures: }
This topic summarizes ``runs but not trustworthy'' FL failures, where experiments execute without crashing but training dynamics or evaluation results are unreliable. Developers report numerical instability (NaN/Inf losses, overflow/underflow, accuracy collapse), often amplified by federated conditions such as non-IID data, low client participation, and sensitive algorithm terms (e.g., FedProx, FedDyn, qFedAvg). Evaluation is also fragile: test metrics may become NaN, stay at zero, or fluctuate unrealistically due to incorrect thresholds, dataset splits, or evaluating the wrong model/client state, especially in object detection. Additional problems come from model--data interface mismatches, checkpoint/pretrained-weight confusion across multiple server/client states, and edge-case bugs in federated loops (e.g., division-by-zero, missing test data). Overall, the topic highlights that small configuration or evaluation mistakes can make FL results unstable or misleading even when the code runs end-to-end.

A telling illustration is the post \textit{``How to use pretrained weights in object detection''}. The developer describes an object-detection workflow where training proceeds normally, and they load pretrained YOLOv5 weights expecting improved convergence and evaluation performance. However, the validation logs exhibit degenerate behavior: mAP remains at zero across epochs or yields implausible values that do not match the apparent training progress. The discussion centers on why the metrics are unreliable despite an end-to-end run, raising issues such as whether evaluation is being performed with the correct thresholds and dataset split, whether the validation loader is aligned with the training preprocessing, and whether the system is evaluating the intended model state (e.g., a stale checkpoint, a client-local model, or an incorrectly restored server/global model). The thread also considers checkpoint and weight-loading pitfalls that can silently bypass the intended pretrained initialization. In this way, the example reflects the core concern of this topic: in FL pipelines, evaluation can appear to execute correctly while still producing misleading results due to subtle state, configuration, or data-handling mismatches.

\textbf{7. FATE Flow Operations, CLI, \& Data-Management: } This topic summarizes This topic summarizes operational and developer-experience issues around FATEFlow, the orchestration layer in the FATE ecosystem. Common problems include starting services, connecting dependencies (especially ZooKeeper), and running jobs through the CLI/SDK. Developers also struggle with data onboarding (upload/table binding, namespaces, schema/SID settings, and storage backends such as MySQL/HDFS/Eggroll/Spark). A major pain point is weak observability: pipelines may run but produce empty or incorrect outputs, and logs or error messages are often insufficient for diagnosing partial failures. Configuration mismatches between standalone and cluster modes (e.g., \texttt{workmode.cluster1}) further increase failure risk. Overall, the topic highlights that reliability, clearer tooling, better documentation, and stronger production features (e.g., HA, authentication, security) are critical as FATE moves from experiments to real deployments.

A clear illustration is the issue titled \textit{``Error Message report incorrectly when task failed in only one party in FATE-1.2''}. The developer describes a multi-party job where one participant fails during execution, yet the surfaced error message is misleading or incomplete, making it difficult to determine which party failed and why. Because the orchestration layer does not clearly propagate party-level failure details to the top-level job status, the overall pipeline can appear to complete while the produced outputs are missing or incorrect. The thread emphasizes that the CLI output and available logs do not provide enough signal to pinpoint the partial failure, forcing users to manually inspect multiple components and party environments to reconstruct what happened. This discussion aligns with the core theme of the topic: limited observability and weak failure reporting in multi-party orchestration can undermine trust in results and significantly increase the operational burden of running FATEFlow pipelines.

\textbf{8. Federated Feature Engineering \& SecureBoost Issues: }
This topic summarizes feature-engineering challenges in FL pipelines (especially binning, one-hot encoding, and feature selection) that are tightly coupled with HeteroSecureBoost-style models. In federated settings, these steps become distributed protocols with privacy constraints and communication overhead, so developers face issues in performance, correctness, usability, and deployment. Common problems include slow binning and high inference latency, large payloads, and robustness bugs such as bin-count mismatches, missing-value handling errors, and merge failures across parties. Developers also request better modularity, logging, and support for multiple feature-selection strategies with safeguards against degenerate cases. Privacy adds complexity because features may need anonymized identifiers, and developers worry about leakage during binning or early splits. Finally, production concerns include online inference support, caching, multi-host serving, and compatibility across homo/hetero variants, alongside dependency-security maintenance. Overall, the topic highlights that federated feature engineering is a major systems challenge that must balance speed, correctness, privacy, and deployability.

One concrete illustration is the issue \textit{``Hetero Feature Binning Miscounts Bin Number for Data with Missing Values''}. The developer reports that when the input data contain missing values, the binning procedure produces an inconsistent number of bins, leading to mismatched binning summaries across parties. The discussion emphasizes that this is not merely a local preprocessing bug: in a federated HeteroSecureBoost pipeline, binning statistics must remain consistent between guest and host despite privacy constraints and distributed execution. When missing values cause bin-count discrepancies, the inconsistency can propagate into downstream training, resulting in runtime errors during merge steps, misaligned feature representations between parties, or silent degradation in model quality due to incorrect split decisions. The thread therefore illustrates a central robustness concern captured by this topic: federated feature engineering must handle abnormal values such as missing entries while preserving cross-party consistency and protocol correctness under communication and privacy constraints.

\textbf{9. FL Capability Gaps Across Settings: } 
This topic summarizes the recurring ``does it support X?'' and ``how do I implement Y?'' questions that arise when developers use FL frameworks in real projects. The discussions reflect uncertainty about feature coverage, availability of runnable examples, and how to integrate frameworks into diverse FL workflows (horizontal, vertical, decentralized, asynchronous) and deployment settings (simulation vs.\ real multi-machine, cloud--edge, multi-server). Developers also ask about compatibility with many model families beyond standard neural networks, including tree-based models, GANs, object detection, graph learning, heterogeneous models, and large-model/LLM-style training. Vertical FL is a major pain point due to role management (guest/host/arbiter), feature alignment, and secure protocol steps, which require strong end-to-end examples. Requests also extend to advanced variants (personalization, robustness), customization of aggregation/optimizers, and end-to-end needs such as evaluation correctness, privacy mechanisms, interpretability, and auditability. Overall, the topic highlights the gap between FL concepts and practical, well-documented, reproducible implementations.

One illustrative sample is titled \textit{``Vertical Federated Learning Examples Help''}. The developer explicitly asks for runnable vertical FL examples beyond the standard FedAvg-style tutorials, emphasizing that high-level descriptions are not sufficient to implement the workflow correctly. The discussion highlights the elements that must be demonstrated concretely: how to assign and manage roles such as guest/host/arbiter, how to perform feature alignment across parties, and how secure protocol steps are orchestrated end-to-end in code. The request also reflects a broader uncertainty about whether vertical FL is supported as a complete, usable pipeline or only described at the conceptual level, since the developer cannot find a reference implementation that they can execute and adapt. In this way, the post matches a central pattern in this topic: when frameworks lack complete, reproducible examples for advanced settings like vertical FL, developers cannot confidently determine feature coverage or implement the intended workflow in practice.

\textbf{10. KubeFATE Deployment \& Operations Troubleshooting: } 
This topic summarizes operational challenges in deploying KubeFATE/FATE, mainly on Kubernetes (and sometimes via \texttt{docker-compose}). The key problems are not algorithmic, but infrastructure-related: deployments fail due to mismatched versions across the KubeFATE CLI/services, Helm charts, Kubernetes, and Helm APIs, often worsened by Kubernetes deprecations (e.g., older \texttt{Ingress} APIs, \texttt{PodSecurityPolicy}). Developers also struggle with networking and access configuration (incorrect \texttt{serviceUrl}, ingress/NodePort/DNS issues), authentication failures (\texttt{401} errors, login/token problems), and Helm chart distribution limits (timeouts, unreachable repos in restricted clusters). Cluster stability issues such as PVC failures, \texttt{CrashLoopBackOff}, and pods stuck initializing further block usability. Limited observability (weak logs/status commands) and platform gaps (e.g., ARM support) amplify debugging difficulty. Overall, the topic highlights that reliable KubeFATE usage depends on careful version alignment, networking, auth, storage, and operational visibility.

One concrete example is the issue \textit{``Error in deploy k8s-deploy with `helm install error' and `go-sqlite3'''}. The developer attempts a Helm-based KubeFATE deployment and encounters an installation failure that references \texttt{go-sqlite3}, indicating a toolchain or host-environment incompatibility rather than an FL algorithm problem. The discussion focuses on how the deployment pipeline spans multiple layers---the KubeFATE CLI/deployer, Helm chart installation, Kubernetes cluster APIs, and host-level build/runtime dependencies---and how a mismatch at any layer can cause the entire installation to fail. In this case, the error suggests that the deployer depends on components that require a compatible Go build environment or native libraries, and the missing or incompatible dependency blocks progress before any services can start. The thread therefore reflects the central theme of this topic: KubeFATE deployments are often constrained by version alignment and infrastructure prerequisites, and failures can arise from the interaction between Helm/Kubernetes operations and lower-level environment dependencies.

\textbf{11. Docker pull/build/run issues: }
This topic summarizes Docker-related deployment failures in FL systems, where workflows break because Docker images, tags, platforms, or \texttt{docker-compose} templates do not match the documentation or the user’s environment. Common issues include missing image tags/manifests (e.g., no \texttt{latest}, changed tag conventions, incomplete multi-arch manifests), architecture mismatches (ARM pulling \texttt{amd64} images leading to \texttt{exec format error}), and unreliable image builds (Dockerfile/BuildKit failures, missing system libraries). Even when containers start, they may exit quickly due to missing Python modules, broken entrypoints, skipped initialization, or missing persistent state (e.g., SQLite tables such as \texttt{t\_job}). Compose files often fail due to version drift in service names, environment variables, image tags, or startup ordering. Overall, the topic shows that Docker is a major friction point when images, build steps, runtime initialization, and orchestration templates are not kept in sync with documentation and platform support.

A telling example is the issue \textit{``docker-compose template for mnist-keras does not work out of the box''}. The developer follows the documented quick-start instructions and runs the provided \texttt{docker-compose} file expecting a runnable MNIST--Keras pipeline, but the deployment fails when executed as-is. The discussion points to orchestration template drift: the compose specification appears to assume specific image tags, service names, environment variables, and startup dependencies that no longer match the current images or application expectations. As a result, containers may fail to pull due to missing tags, start and immediately exit due to missing modules or broken entrypoints, or come up in an invalid order that leaves dependent services uninitialized. The thread highlights that even when users follow the recommended path exactly, they can be blocked because the compose template, referenced Docker images, and runtime initialization steps have fallen out of sync with the evolving codebase and documentation.

\textbf{12. GPU device \& memory misconfiguration: }
This topic summarizes GPU-related failures in FL workloads, especially in multi-GPU settings. Developers report that training either crashes or runs inefficiently because GPU visibility, device mapping, and configuration do not align with real execution environments (e.g., HPC/Slurm, VMs, mixed CPU/GPU setups). Common problems include incorrect GPU selection (workloads collapsing onto \texttt{GPU0} due to hard-coded indices or mismanaged \texttt{CUDA\_VISIBLE\_DEVICES}), invalid device indexing (\texttt{invalid device ordinal}), and GPU memory issues (unit mismatches and memory growth leading to OOM across rounds). Additional fragility comes from mixed CPU/GPU tensor handling and checkpoint loading (often requiring \texttt{map\_location=`cpu`}). Default configuration assumptions (e.g., missing parameters like \texttt{nproc\_per\_node} or inability to run CPU-only clients) further break deployments. Overall, reliable GPU use in FL requires consistent device visibility, indexing, memory accounting, and configuration semantics across distributed components; otherwise training may fail or silently use unintended devices.

One illustrative thread is titled \textit{``Offsite tuning code with multigpu setting throws error''}. The developer runs an FL workload in an HPC/Slurm-style environment and configures the framework to use multiple GPUs, expecting different clients or processes to be mapped across available devices. The logs suggest that the system recognizes multiple GPUs and reports assigning clients to distinct GPU groups, yet the observed runtime behavior indicates that execution still collapses onto \texttt{GPU0} (or fails with device-indexing errors). The discussion points to inconsistent enforcement of device visibility and mapping across layers: environment variables such as \texttt{CUDA\_VISIBLE\_DEVICES} may not be applied uniformly across spawned processes, the framework's scheduling logic may reference physical device indices that do not match the visible ordering inside each job step, and the underlying deep-learning runtime may default to \texttt{cuda:0} when device selection is not propagated correctly. As a result, the intended multi-GPU configuration does not translate into actual multi-GPU execution, illustrating the central problem in this topic: without consistent GPU visibility and indexing semantics, FL workloads can silently use unintended devices or fail unpredictably in distributed environments.

\textbf{13. Runtime \& RPC Failures} 
This topic summarizes post-startup failures in FL platforms: the system launches successfully, but later breaks during job submission, scheduling, component execution, or cross-party communication. Developers report opaque tracebacks and non-zero return codes (e.g., \texttt{retcode 103/104/105}) from orchestration and runtime layers. Common issues include job-creation/scheduling failures (orchestrator cannot coordinate roles or reach backend services), data onboarding and table/namespace errors (uploads fail or datasets cannot be found due to inconsistent registration), and communication instability (gRPC timeouts, deadline exceeded, unsupported HTTP operations caused by endpoint/port/proxy or version mismatches). Additional failures involve serialization/type-handling problems (non-JSON-serializable objects, schema mismatches) and resource/process-management issues (defunct processes, unstable worker pools, leaked resources) that appear under repeated or long-running execution. Overall, the topic highlights that successful startup is not enough; dependable FL requires robust orchestration, communication, serialization, and lifecycle management.

An instance that can be found is the post \textit{``TypeError: The supplied argument maps to TFF type \ldots which is incompatible with the requested type''}. The developer describes a workflow that appears to initialize correctly: the environment starts, an iterative federated process is constructed, and the code reaches the point of executing a training round. Execution then fails mid-run with a long traceback when TFF enforces strict type signatures and detects that the supplied client-side structure does not match the expected federated type (e.g., a mismatch in dataset element structure, nesting, or \texttt{TensorSpec} fields). The discussion emphasizes that this kind of failure can be difficult to diagnose because the system does not crash at startup, but instead surfaces the incompatibility only when the computation is invoked and type-checking is applied to the placed values. As a result, the developer must trace back through the data pipeline and model \texttt{input\_spec} to find the subtle representation mismatch. This thread aligns with the core theme of the topic: post-startup failures often arise from orchestration and runtime enforcement layers, producing opaque exceptions that make debugging challenging even when the workflow initially appears to run.
\newline
\begin{center}
\setlength{\fboxsep}{10pt}
\fbox{
\parbox{0.94\linewidth}{
\textbf{RQ1 Summary:}
Stack Overflow discussions primarily reflect user-facing challenges related to
framework implementation, installation, data preparation, model-state
management, and evaluation. GitHub artifacts more strongly capture
repository-level concerns involving configuration, reproducibility,
infrastructure, maintenance, and deployment. Together, the two platforms
provide complementary views of FL development, ranging from learning and
experimentation to system operation and evolution.
}}
\end{center}

\subsection{RQ2: What types of FL development questions are asked on technical forums?}
\textbf{Motivation: } Building upon identifying key topics in \textbf{RQ1}, a critical question emerges: \emph{What kinds} of questions are FL developers asking across community platforms? Answering this matters because it relies on understanding the role that both SO and GitHub play in supporting FL developers. This investigation shifts the focus from only what developers ask to also why they ask it. By analyzing the types of content created within each FL topic—such as SO questions and GitHub issues/discussion threads—we aim to uncover the nature of the challenges developers face during FL development. Prior research \cite{rosen2016mobile, abdellatif2020challenges} shows that developers usually ask different types of questions—such as “How”, “Why”, and “What”—to deal with specific difficulties. This detailed analysis will not only surface the most frequent problems, but also reveal the deeper complexities and pain points in the FL development process. Categorizing these questions and reports will provide insight into developers’ thinking and problem-solving approaches, helping identify areas where more guidance, tooling, or learning resources are needed. Ultimately, this work offers a deeper understanding of why FL developers turn to SO and GitHub for help instead of relying only on official documentation.

\textbf{Approach: } To identify the types of FL-related posts, we used the same high-level taxonomy commonly adopted in prior work (i.e., \emph{How}, \emph{What}, and \emph{Why}) \cite{rosen2016mobile, abdellatif2020challenges, treude2011programmers}. Instead of just manually labeling a sample, we analyzed the \emph{full} dataset, which includes \textbf{495 Stack SO posts} and \textbf{9,116 GitHub issues} collected independently for the FL topics identified in \textbf{RQ1}. We chose this strategy because GitHub contains thousands of issues, and fully manual labeling would be time-consuming, hard to reproduce consistently and may not be the most robust way.

To label post types at scale, we implemented a \textbf{hybrid automatic classification pipeline} (see the provided Python script in replication package\citep{saed2026flchallenges}). For each SO post/GitHub issue, we used the available text field (e.g., title/body combined in the \texttt{Text} column) and assigned one of four labels: \emph{how}, \emph{what}, \emph{why}, or \emph{other}. The pipeline works in three steps:

(1) \textbf{Rule-based labeling (high precision).} We first apply simple and interpretable regex rules to detect explicit signals of each type. For example, if the text starts with “how/what/why”, or contains strong indicators such as troubleshooting terms (e.g., \emph{error}, \emph{exception}, \emph{not working}), it receives a score for the corresponding class. If the top class is sufficiently strong and clearly higher than the runner-up (using the script thresholds), we directly assign that label.

(2) \textbf{Zero-shot labeling for unclear cases (better coverage).} If the rule-based step is not confident (i.e., ambiguous wording or weak signals), the post is sent to a zero-shot classifier (\texttt{facebook/bart-large-mnli}) with three candidate intents: \emph{instructions/troubleshooting (how)}, \emph{definition/explanation (what)}, and \emph{reason/cause (why)}. If the model confidence is low or the difference between the top two classes is small, we label the post as \emph{other} to avoid forcing an unreliable type.

(3) \textbf{Hybrid decision + aggregation.} We keep the rule-based prediction whenever it is confident; otherwise, we use the zero-shot prediction (or \emph{other} when uncertain). Finally, we compute the percentage distribution of types per topic for both platforms.

\paragraph{Manual validation of the question-type classification.}
To validate the hybrid classification pipeline, we manually labelled a
stratified sample of 440 artifacts, comprising 180 Stack Overflow posts
and 260 GitHub issues and pull requests. We sampled 20 artifacts from
each of the nine Stack Overflow topics and thirteen GitHub topics,
ensuring coverage of both platforms and all identified topics.

The manual classification was conducted jointly by two authors: one
with more than two years of FL research experience and another with more
than five years of empirical software engineering research experience.
The authors discussed each artifact and assigned a consensus label of
How, Why, What, or Other. For ambiguous cases, they consulted a third
author with more than ten years of software engineering research
experience.

We then compared the consensus manual labels with those produced by the
hybrid rule-based and zero-shot classification pipeline. Agreement
between the manual and automatic classifications was measured using
Cohen's kappa and reached $\kappa=0.78$, indicating substantial
agreement. This result provides additional confidence in applying the
automatic classification pipeline to the complete dataset.

\vspace{2mm}

\begin{itemize}
    \item \textbf{How: }This category includes posts where developers seek guidance on specific methods or techniques to accomplish a particular task \cite{rosen2016mobile, abdellatif2020challenges}. Unlike 'Why' posts that inquire about underlying reasons or explanations, these posts are goal-oriented. Developers ask for precise, step-by-step instructions to achieve their specific objectives \emph{(e.g., 'How to remove default noise in FL frameworks?')}.
     \item \textbf{Why: }This type includes posts where developers inquire about certain phenomena' reasons, causes, or purposes \cite{rosen2016mobile, abdellatif2020challenges}. 'Why' posts often pertain to troubleshooting, where developers seek explanations for specific behaviors or issues \emph{(e.g., 'Why does aggregating need a message broker?')}.
    \item  \textbf{What: }This category includes posts where developers seek specific information \cite{rosen2016mobile, abdellatif2020challenges}. These inquiries are often aimed at clarifying doubts to make more informed decisions. For example, a developer asked, \emph{'What is the best approach for adding cron jobs (scheduled tasks) for a particular service in Docker Compose?'} in SO.
    \item \textbf{Others: }We assigned this type to posts that do not fit into any of the above-described types \emph{(e.g., 'Understand the Notify and Wait process in NIFI in my flow?')}.
\end{itemize}

\renewcommand{\arraystretch}{1.5}
\begin{table}[htbp]
    \centering
    \caption{FL posts types distribution on Stack Overflow}
    \label{tab:FLSOtopicstypes}
    \begin{tabular}{l|c|r|r|r}
        \toprule
        Topic & \% How & \% Why & \% What & \% Other \\
        \midrule
        TFF Implementation \& Client & 43.71\% & 14.37\% & 33.53\% & 8.33\% \\
        TFF API \& Compatibility Errors & 68.42\% & 10.52\% & 13.15\% & 7.84\% \\
        FL Convergence \& Evaluation Mismatch & 41.81\% & 21.81\% & 21.81\% & 14.45\% \\
        TFF Model State \& Aggregation & 45.28\% & 15.09\% & 33.96\% & 5.66\% \\
        TFF Installation \& Environment Compatibility & 83.33\% & 6.25\% & 6.25\% & 4.16\% \\
        Data Partitioning and Loading for FL &  50\% & 15\% & 30\% & 5\% \\
        PySyft Setup \& API Breakages & 66.66\% & 23.80\% & 4.76\% & 4.76\% \\
        DP in FL (Clipping \& Noise) & 20\% & 15\% & 60\% & 5\% \\
        FL Frameworks – Scope \& Deployment Reality & 20\% & 13.33\% & 53.33\% & 13.33\% \\
        \midrule
        FL (all) & 51.11\% & 14.34\% & 26.66\% & 7.87\% \\
        \bottomrule
    \end{tabular}
\end{table}

\renewcommand{\arraystretch}{1.5}

\renewcommand{\arraystretch}{2}
\begin{table}[htbp]
    \centering
    \caption{FL issue types distribution on GitHub}
    \label{tab:FLGHtopicstypes}
    \begin{tabular}{p{0.5\textwidth}|c|r|r|r}
        \toprule
        Topic & \% How & \% Why & \% What & \% Other \\
        \midrule
        FL System Deployment and Operational Failures & 27.78\% & 41.96\% & 19.17\% & 11.11\% \\
        Federated Data Partitioning \& Pipeline & 13.2\% & 50.22\% & 23.75\% & 12.43\% \\
        FL Debugging, Configuration, \& Operations & 5.37\% & 77.97\% & 8.63\% & 8.03\% \\
        FL Benchmarks Reproducibility Issues & 18.04\% & 49.54\% & 18.34\% & 14.06\% \\
        Federated Wiki Docs, Navigation \& Collaborations & 19.67\% & 37.74\% & 28.06\% & 14.51\% \\
        Training instability and evaluation failures & 38.98\% & 36.88\% & 14.83\% & 9.32\% \\
        FATE Flow Operations, CLI, and Data-Management & 18.43\% & 60.36\% & 16.12\% & 5.09\% \\
        Federated Feature Engineering and SecureBoost Issues & 12.29\% & 39.10\% & 41.34\% & 7.26\%\\
        FL Capability Gaps Across Settings & 10.97\% & 46.04\% & 32.37\% & 10.79\% \\
        KubeFATE Deployment \& Operations Troubleshooting & 45.34\% & 33.72\% & 12.79\% & 8.13\% \\
        Docker pull/build/run issues & 31.64\% & 45.56\% & 11.39\% & 11.39\% \\
        GPU device \& memory misconfiguration & 49.29\% & 36.66\% & 9.85\% & 4.22\% \\
        Runtime \& RPC Failures & 49.29\% & 25.35\% & 15.49\% & 9.85\% \\
        \midrule
        FL (all) & 25.76\% & 43.85\% & 19.46\% & 10.91\% \\
        \bottomrule
    \end{tabular}
\end{table}
\renewcommand{\arraystretch}{2}

Although the How--Why--What taxonomy was originally developed for
question-oriented discussions, we use it here as a common abstraction
of the dominant communicative intent expressed across both platforms.
GitHub issues and pull requests may also represent bug reports, feature
requests, maintenance tasks, or coordination activities rather than
explicit questions. Artifacts without a clear How, Why, or What intent
were therefore classified as \emph{Other}.

\paragraph{Cross-platform statistical comparison.}
To quantify differences in question intent between Stack Overflow and
GitHub, we constructed a $2 \times 4$ contingency table containing
platform and question type. The How, Why, What, and Other categories
were defined consistently across both platforms and therefore provided
a common basis for direct statistical comparison.

We applied Pearson's chi-square test of independence to examine the
association between platform and question type and calculated Cramér's
$V$ to quantify the overall effect size. We also calculated adjusted
standardized residuals to identify the question types contributing most
strongly to the observed association. For each question type, we
calculated an odds ratio with a 95\% confidence interval by comparing
that category with the remaining three categories. Holm's procedure was
applied to the four category-level $p$-values to account for multiple
comparisons. All analyses were conducted using artifact-level counts
rather than rounded percentages.

\textbf{Results: }Tables~\ref{tab:FLSOtopicstypes} and~\ref{tab:FLGHtopicstypes} summarize the percentage of post types (\emph{How}, \emph{What}, \emph{Why}, and \emph{Other}) for each FL topic on Stack Overflow (SO) and GitHub, respectively. The distributions show clear patterns that reflect how FL developers seek help on different platforms.

On \textbf{Stack Overflow} (495 posts), the most common type is \textbf{\emph{How}}, with an average of \textbf{48.80\%} across topics. This indicates that FL developers mainly use SO to request procedural guidance and troubleshooting support (e.g., how to implement, configure, or fix issues). Some topics are particularly dominated by \emph{How} posts, such as \textbf{TFF Installation \& Environment Compatibility (83.33\%)}, followed by \textbf{TFF API \& Compatibility Errors (68.42\%)} and \textbf{PySyft Setup \& API Breakages (66.67\%)}. In contrast, \textbf{\emph{What}} posts represent \textbf{28.54\%} on average and become the largest category in some topics, most notably \textbf{DP in FL (Clipping \& Noise) (60.00\%)} and \textbf{FL Frameworks – Scope \& Deployment Reality (53.33\%)}, suggesting a strong need for clarifications of FL concepts, APIs, and expected behaviors. \textbf{\emph{Why}} posts are less frequent overall (\textbf{15.02\%} on average), although they still appear in noticeable proportions in topics such as \textbf{PySyft Setup \& API Breakages (23.81\%)} and \textbf{FL Convergence \& Evaluation Mismatch (21.82\%)}. Finally, the \textbf{\emph{Other}} category remains relatively small (\textbf{7.64\%} on average), meaning most SO posts can be captured well by the \emph{How/What/Why} taxonomy.
On \textbf{GitHub} (9,116 issues), issue discussions are more often framed as \emph{explanations of behavior} than as direct ``how-to'' questions. Across topics, \textbf{\emph{Why}} is the most common intent (mean \textbf{44.72\%}), meaning that many issues ask for a reason or root cause---for example, why a framework behaves unexpectedly, why results change between runs, or why an error occurs. This pattern is strongest in \textbf{FL Framework Debugging, Configuration, \& Operations} (\textbf{77.98\%}) and remains high in \textbf{FATE Flow Operations, CLI, \& Data-Management} (\textbf{60.37\%}).

In contrast, \textbf{\emph{How}} intents (mean \textbf{26.17\%}) reflect fix-oriented requests, where developers ask for concrete steps to make a system work. \textbf{\emph{How}} dominates in several operational topics, including \textbf{GPU device \& memory misconfiguration} (\textbf{49.30\%}), \textbf{Runtime \& RPC Failures} (\textbf{49.30\%}), and \textbf{KubeFATE Deployment \& Operations Troubleshooting} (\textbf{45.35\%}), highlighting a strong need for reproducible troubleshooting guidance. \textbf{\emph{What}} intents account for a smaller but meaningful share (mean \textbf{19.40\%}) and are most frequent in topics where developers seek clarification about capabilities or concepts, such as \textbf{Federated Feature Engineering \& SecureBoost Issues} (\textbf{41.34\%}) and \textbf{FL Capability Gaps Across Settings} (\textbf{32.37\%}). Finally, \textbf{\emph{Other}} remains limited (mean \textbf{9.71\%}) but is slightly higher than on Stack Overflow, which is expected because GitHub threads also include feature requests, maintenance planning, and coordination discussions that do not fit a strict Q\&A framing.

Similar to what prior studies have observed in other developer-support contexts, \emph{How}-type posts appear frequently because developers often seek direct, actionable steps to resolve implementation problems \cite{rosen2016mobile, abdellatif2020challenges, treude2011programmers}. However, FL shows an interesting difference on SO: \textbf{\emph{What} questions are more common than \emph{Why} questions} (28.54\% vs. 15.02\%). This likely reflects the fact that FL involves many new concepts, components, and configuration choices, so developers often first need to understand \emph{what} something is or \emph{what} a specific option/behavior means before they can effectively implement or debug it.

Overall, the tables highlight a strong \textbf{platform difference} in how FL developers seek support: \textbf{SO is mainly “How” oriented} (instruction and troubleshooting), while \textbf{GitHub is mainly “Why” oriented} (reasoning and root-cause discussions). This suggests that FL developers rely on SO for quick, implementation-focused assistance, and turn to GitHub when the problem is tied to a specific library/version or requires deeper explanation and diagnosis.

\paragraph{Quantitative cross-platform comparison.}
The question-type distribution differed significantly between Stack
Overflow and GitHub, $\chi^{2}(3, N=9{,}611)=224.30$, $p<0.001$.
Cramér's $V$ was 0.153, indicating a small overall association between
platform and question intent.

As shown in Table~\ref{tab:cross_platform_intent}, How-type artifacts
accounted for 51.11\% of Stack Overflow posts, compared with 25.77\%
of GitHub artifacts. The odds of an artifact being classified as How
were 3.01 times greater on Stack Overflow than on GitHub
($OR=3.01$, 95\% CI: 2.51--3.61). What-type artifacts were also more
prevalent on Stack Overflow than on GitHub, at 26.67\% and 19.46\%,
respectively ($OR=1.50$, 95\% CI: 1.22--1.85).

In contrast, Why-type artifacts accounted for 43.86\% of GitHub
artifacts but only 14.34\% of Stack Overflow posts. The
Stack Overflow-to-GitHub odds ratio was 0.21
(95\% CI: 0.17--0.28), which corresponds to approximately 4.66 times
greater odds of a Why-type classification on GitHub. Other-type
artifacts were also more prevalent on GitHub, at 10.91\%, compared
with 7.88\% on Stack Overflow. The corresponding Stack
Overflow-to-GitHub odds ratio was 0.70
(95\% CI: 0.50--0.97), equivalent to approximately 1.43 times greater
odds on GitHub.

The adjusted standardized residuals further showed that How-type and
What-type artifacts were overrepresented on Stack Overflow, with
residuals of 12.36 and 3.92, respectively. Why-type and Other-type
artifacts were underrepresented on Stack Overflow, with residuals of
$-12.94$ and $-2.12$, respectively. The corresponding GitHub
residuals had the same magnitudes in the opposite directions. All four
category-level comparisons remained statistically significant after
Holm adjustment.

These findings quantitatively support the distinct roles of the two
platforms. Stack Overflow is more strongly associated with procedural
guidance, implementation, and information-seeking questions, whereas
GitHub is more strongly associated with explanation, diagnosis, and
investigation of framework- and project-specific behaviour.

\begin{table}[htbp]
\centering
\caption{Quantitative comparison of question types across Stack
Overflow and GitHub. Odds ratios compare Stack Overflow with GitHub;
values greater than one indicate a stronger association with Stack
Overflow.}
\label{tab:cross_platform_intent}

\small
\renewcommand{\arraystretch}{1.15}
\setlength{\tabcolsep}{3pt}

\begin{tabular*}{\columnwidth}{
@{\extracolsep{\fill}}
lrrrr
@{}
}
\toprule
\textbf{Type} &
\textbf{SO, $n$ (\%)} &
\textbf{GitHub, $n$ (\%)} &
\textbf{OR (95\% CI)} &
\textbf{Adj. $p$} \\
\midrule
How
& 253 (51.11\%)
& 2,349 (25.77\%)
& 3.01 (2.51--3.61)
& $<0.001$ \\

Why
& 71 (14.34\%)
& 3,998 (43.86\%)
& 0.21 (0.17--0.28)
& $<0.001$ \\

What
& 132 (26.67\%)
& 1,774 (19.46\%)
& 1.50 (1.22--1.85)
& $<0.001$ \\

Other
& 39 (7.88\%)
& 995 (10.91\%)
& 0.70 (0.50--0.97)
& 0.035 \\
\bottomrule
\end{tabular*}
\end{table}

The odds ratios use GitHub as the reference platform. Therefore, values
greater than one indicate that a question type is more strongly
associated with Stack Overflow, whereas values below one indicate a
stronger association with GitHub. We restrict the inferential
cross-platform analysis to question intent because the same taxonomy
was applied to both datasets. The Stack Overflow and GitHub topic
models were generated independently, and their topic categories do not
form one-to-one equivalent groups. Moreover, accepted Stack Overflow
answers and closed GitHub artifacts represent different resolution
mechanisms and are therefore examined separately in \textbf{RQ3}.
\newline
\begin{center}
\setlength{\fboxsep}{10pt}
\fbox{
\parbox{0.94\linewidth}{
\textbf{RQ2 Summary.}
How-type questions dominate Stack Overflow, indicating strong demand for
procedural guidance, implementation examples, and troubleshooting support.
In contrast, Why-type questions are prominent across several GitHub topics,
where developers seek explanations for unexpected configuration, training,
evaluation, and runtime behaviour. These patterns reflect the different
support roles of the two platforms.
}}
\end{center}

\subsection{RQ3: Topic Difficulty}
\textbf{Motivation: } After identifying the main FL topics from Stack Overflow and GitHub, the next step is to see how hard it is to answer questions and resolve issues and pull requests in each topic. If some topics take longer or remain unresolved more often, this suggests they are harder and may need more community help. This analysis can also point to topics where better tools or frameworks are needed to support FL developers.
\newline
\textbf{Approach: } We assessed topic-level resolution outcomes using two platform-specific metrics. For Stack Overflow, we measured (1) the percentage of posts in each topic without an accepted answer and (2) the median time, in hours, until an answer was accepted. For GitHub, we measured (1) the percentage of issues and pull requests that remained unresolved and (2) their median resolution time. This approach is consistent with methods used in previous studies \citep{rosen2016mobile, bagherzadeh2019going, yang2016security, li2021understanding, scoccia2021challenges, abdellatif2020challenges}. However, these metrics are treated as indirect indicators of observed resolution difficulty rather than direct measures of technical complexity. Stack Overflow outcomes may be influenced by community participation, answer visibility, and user acceptance behaviour, while GitHub outcomes may also reflect maintainer availability, backlog priorities, release planning, and project-management practices. The metrics are described in detail below:
\newline
\textbf{Metrics for SO: }
\begin{enumerate}
    \item \textbf{The percentage of posts of a topic without accepted answers (\% w/o accepted answers).} For each FL topic, we calculated the share of posts that do not have an accepted answer. Even if a question gets many replies, only the asker can accept one answer when it fully solves the problem. So, topics with a higher percentage of unaccepted posts are treated as harder\cite{abdellatif2020challenges, rosen2016mobile, bagherzadeh2019going}.
    
    \item \textbf{The median time for an answer to be accepted (median time to answer (hrs.)).} We also calculated the median time (in hours) it takes for a post to get an accepted answer. We measured this using the creation time of the accepted answer, not the time it was later marked as accepted. If the median time is longer, it suggests the posts in that topic are more difficult\cite{abdellatif2020challenges, rosen2016mobile,bagherzadeh2019going}.
\end{enumerate}

\textbf{Metrics for GitHub: }
\begin{enumerate}
\item \textbf{The percentage of unresolved issues and pull requests of a topic (\% w/o solutions).} The unresolved issues and pull requests percentage within a specific topic shows how many remain open. In software development, closing an issue or pull request usually means the problem has been fixed or the feature has been implemented. Therefore, topics with a higher number of open issues or pull requests are considered more difficult.

\item \textbf{The median time for an issue or pull request to be resolved (Median Time to Resolve(hrs.)).} We calculated the median resolution time, in hours, as the time between when an issue or pull request was created and when it was closed. Only closed items were included because open ones do not have a closing time. A longer resolution time indicates that the issue or pull request is more challenging to resolve.
\end{enumerate}

\renewcommand{\arraystretch}{1.4}
\begin{table}[htbp]
    \centering
    \caption{The difficulty per topic on Stack Overflow data}
    \label{tab:sotopicdifficulty}
    \begin{tabular}{p{0.5\textwidth}|r|r}
        \toprule
        Topic & Posts w/o Accepted (\%) & Median Time (h) \\
        \midrule
        TFF Implementation \& Client &\hfill 69.87\% & 17.55 \\
        TFF API \& Compatibility Errors &\hfill 66.21\% & 10.17 \\
        FL Convergence \& Evaluation Mismatch &\hfill 76.36\% & 17.42 \\
        TFF Model State \& Aggregation &\hfill 66.03\% & 39.24 \\
        TFF Installation \& Environment Compatibility &\hfill 82.22\% & 12.49 \\
        Data Partitioning and Loading for FL &\hfill 61.53\% & 16.65 \\
        PySyft Setup \& API Breakages &\hfill 76.19\% & 99.19\\
        DP in FL (Clipping \& Noise) &\hfill 40\% & 47.13 \\
        FL Frameworks - Scope \& Deployment Reality &\hfill 53.33\% & 55.63 \\
        \bottomrule
    \end{tabular}
\end{table}
\renewcommand{\arraystretch}{1.4}

\textbf{Results: }Table \ref{tab:sotopicdifficulty} presents the percentage of questions without accepted answers and the median time (in hours) to receive an accepted answer for each identified topic of SO data. The topic of \emph{TFF Installation \& Environment Compatibility} has the largest share of unanswered questions while PySyft Setup \& API Breakages  takes the longest time to receive an answer. To understand why so many posts lack accepted answers (82.22\%), we reviewed the posts in this topic. We found that many unanswered posts had low scores on Stack Overflow, with 22 posts receiving a score of zero. This suggests that unclear or poorly written questions may reduce the chances of receiving an accepted answer. Additionally, we explored why the PySyft Setup \& API topic has a significantly higher median response time. We found that questions in this topic often involve complex tasks such as installing and configuring PySyft, managing dependencies and environment compatibility, understanding API changes across versions, and integrating PySyft with existing machine learning workflows. These tasks require careful configuration, troubleshooting, and familiarity with PySyft’s evolving APIs, which significantly increases the time needed to provide an accepted answer.

We also found that \emph{TFF Implementation \& Client}, \emph{FL Convergence \& Evaluation Mismatch}, and \emph{TFF Model State \& Aggregation} exhibit low acceptance rates and high median times to receive an accepted answer. An examination of the data shows that between 66\% and 76\% of posts in these topics do not have accepted answers, with median resolution times ranging from 17.42 to 39.24 hours. These topics typically involve complex and context-dependent challenges, such as framework configuration, model training instability, and environment-specific execution issues. As a result, reproducing these problems without access to a similar setup is difficult, which complicates diagnosis and solution development. Providing a meaningful answer often requires in-depth knowledge of specific FL frameworks, configurations, and experimental conditions, significantly increasing the time required to resolve these questions.

Conversely, posts related to \emph{DP in FL (Clipping \& Noise) } and \emph{FL Frameworks - Scope \& Deployment Reality} show comparatively higher acceptance rates and lower median resolution times. In particular, \emph{DP in FL (Clipping \& Noise)} has the lowest proportion of posts without accepted answers (40\%), while \emph{FL Frameworks - Scope \& Deployment Reality} maintains a moderate acceptance rate with a median response time of 55.63 hours. These topics generally involve more straightforward implementation questions or clearer problem statements, making them easier for Stack Overflow contributors to address. Their closer alignment with standard software development and machine learning practices likely explains why accepted answers are provided more frequently and efficiently in these cases.

\renewcommand{\arraystretch}{1.5}
\begin{table}[htbp]
    \centering
    \caption{The difficulty per topic on on GitHub}
    \label{tab:unresolved_issues_median_time}
    \begin{tabularx}{\linewidth}{X|r|r}
        \toprule
        \textbf{Topic Name} & \textbf{Unresolved (\%)} & \textbf{Median Time (h)} \\
        \midrule
        FL Deployment and Operational Failures & 14.83\% & 525.36  \\
        Federated Data Partitioning \& Pipelines & 21.54\% & 277.93 \\
        Debugging, Configuration, \& Operations & 17.55\% & 616.44 \\
        FL Benchmark Reproducibility Issues & 19.87\% & 192.37\\\
        Federated Wiki Docs, Navigation \& Collaboration & 35.80\% & 161.81 \\
        Training instability and evaluation failures & 21.18\% & 136.25  \\
        FATE Flow Operations, CLI, and Data-Management & 3.22\% & 1029.63  \\
        Federated Feature Engineering and SecureBoost Issues & 0.0\% & 1652.78 \\
        FL Capability Gaps Across Settings & 24.46\% & 879.29 \\
        KubeFATE Deployment \& Operations Troubleshooting  & 23.25\% & 521.27 \\
        Docker pull/build/run issues & 20.25\% & 837.71 \\
        GPU device \& memory misconfiguration & 26.76\% & 92.83 \\
        Runtime \& RPC Failures & 4.22\% & 6491.59\\
        \bottomrule
    \end{tabularx}
\end{table}
\renewcommand{\arraystretch}{1.5}
Table~\ref{tab:unresolved_issues_median_time} presents the proportion of unresolved issues and the median time to resolve issues across different FL-related topics based on GitHub issue and pull request data. This analysis highlights persistent developer challenges and reveals how the difficulty of resolving issues varies across topics in FL development.

Notably, \emph{Documentation, Wiki Navigation, and Collaboration Semantics in
Federated Wiki} exhibits the highest unresolved rate (35.81\%) alongside a relatively high median resolution time (161.81 hours), indicating substantial difficulty in addressing issues within this topic. Similarly, \emph{Federated Dataset Engineering and Partitioning Pipeline Challenges} (21.55\% unresolved, 277.93 hours), \emph{FL Capability and Implementation Gaps Across Paradigms and Deployment Settings} (24.46\% unresolved, 879.30 hours), and \emph{KubeFATE Deployment and Operations Troubleshooting on Kubernetes and Docker Compose} (23.26\% unresolved, 521.28 hours) also demonstrate high unresolved proportions coupled with long resolution times. These patterns suggest that topics associated with complex FL configurations, framework interactions, or advanced training workflows pose significant challenges for developers.

In contrast, \emph{Federated Feature Engineering and SecureBoost Issues} shows no unresolved issues (0\%) despite having a very high median resolution time (1652.78 hours), indicating that while issues in this topic are eventually resolved, they require extensive time and effort. Likewise, \emph{FATE Flow Operations, CLI, and Data-Management} (3.23\% unresolved, 1029.64 hours) and \emph{Runtime \& RPC Failures } (4.23\% unresolved, 5491.60 hours) display very low unresolved rates but exceptionally long median resolution times, reflecting the complex and time-consuming nature of these issues rather than a lack of resolution.

Overall, the results demonstrate that unresolved rates and resolution times vary considerably across FL topics. Topics with high unresolved proportions point to areas where developers may struggle to obtain sufficient support, while topics with low unresolved rates but extremely long resolution times suggest deeply complex problems that require significant effort to resolve. These findings emphasize the need for improved tooling, clearer documentation, and stronger community support to reduce recurring challenges in FL development.

\begin{center}
\setlength{\fboxsep}{6pt}
\fbox{
\parbox{0.94\linewidth}{
\textbf{RQ3 Summary.}
Resolution difficulty varies substantially across FL topics. High proportions
of posts without accepted answers or unresolved GitHub artifacts indicate
areas where adequate support may be unavailable. In contrast, low unresolved
rates combined with long resolution times identify complex engineering
problems that are eventually addressed but require substantial diagnosis,
coordination, and maintenance effort.
}}
\end{center}

\section{Discussion \& Implications of our Findings}\label{discussion-implications}

\subsection{FL Topics Evolution}
Federated learning (FL) has rapidly transitioned from a research concept to a widely adopted development paradigm, motivated by the need to train models without centralizing sensitive or regulated data \cite{mcmahan2017communication,kairouz2021advances}. This growth is reflected in the expanding ecosystem of open-source and production-grade FL frameworks and platforms, spanning research prototyping and large-scale deployments \cite{bonawitz2019towards,beutel2020flower,he2020fedml,liu2021fate,reina2021openfl}. As FL continues to spread across application domains such as digital health \cite{rieke2020digitalhealth}, developer activity and community discussions have increased accordingly. 

\begin{figure}[htbp]
  \includegraphics[width=\textwidth]{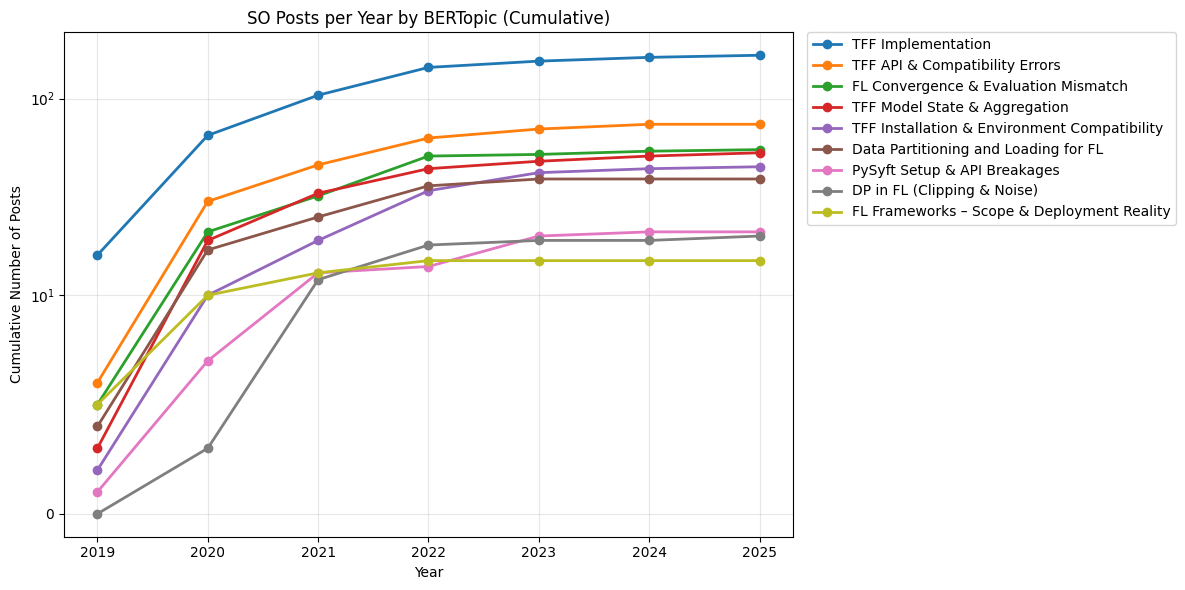}
  \caption{FL topics evolution over time on Stack Overflow}
  \label{fig:SO_Chart}
\end{figure}

\begin{figure}[htbp]
  \includegraphics[width=\textwidth]{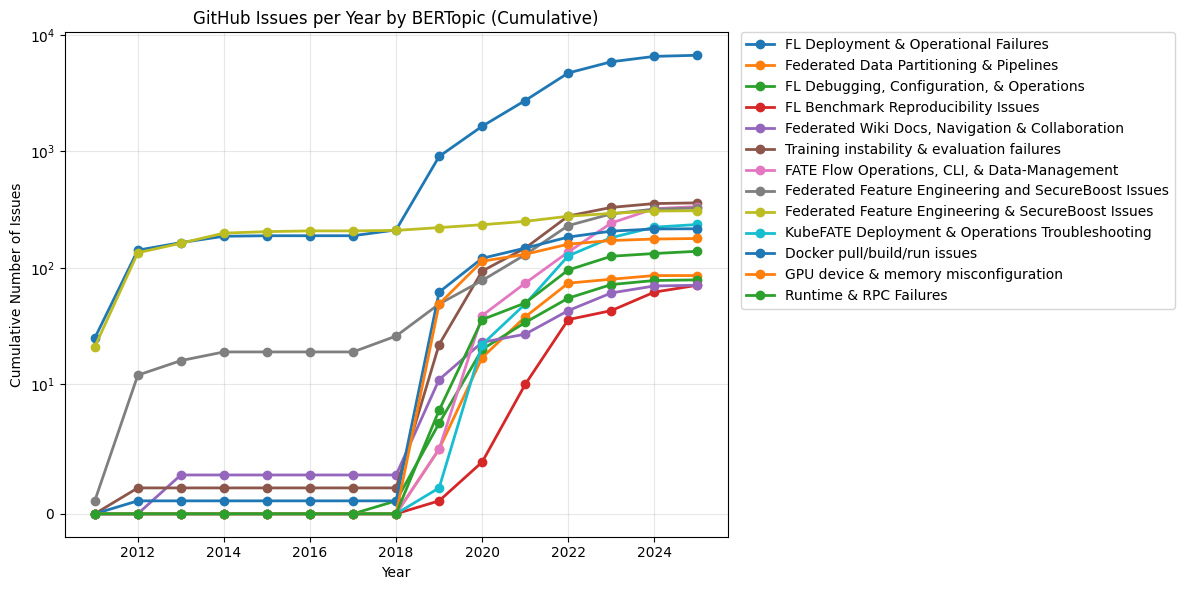}
  \caption{FL topics evolution over time on GitHub}
  \label{fig:GH_Chart}
  \vspace{-1.5em}
\end{figure}

To understand how FL developer challenges evolve over time, we analyzed the year-wise distribution of the identified topics across Stack Overflow (Fig.~\ref{fig:SO_Chart}) and GitHub (Fig.~\ref{fig:GH_Chart}). On Stack Overflow, earlier years were dominated by foundational questions about FL concepts and basic implementation, including algorithm selection and correctness, client sampling, non-IID data behavior, and the mechanics of wrapping local training into a federated loop. Over time, these discussions increasingly shifted toward practical integration concerns such as data pipeline engineering, model interface constraints, and debugging framework-specific execution rules, reflecting the gap between FL theory and the realities of running FL with real models and real datasets.

In parallel, GitHub exhibited a complementary evolution that emphasized implementation, deployment, and maintenance. Persistent topics included framework debugging and configuration, runtime and RPC failures, and infrastructure-level issues such as Docker build and pull failures, GPU device and memory misconfiguration, and scalability limitations under large client populations. The prominence of these themes aligns with the systems challenges highlighted in prior work on production-scale FL, where orchestration, scheduling, and reliability become dominant concerns as FL moves from simulation to deployment \cite{bonawitz2019towards}. Documentation and usability discussions also increased over time, signaling growing attention to onboarding, reproducibility, and operational readiness within the FL tooling ecosystem.

Overall, Stack Overflow discussions more often focus on conceptual understanding, best practices, and troubleshooting common development errors, whereas GitHub issues capture deeper framework-level debugging, infrastructure dependencies, and long-term maintenance. The alignment and divergence of topic trajectories across these platforms reveal the layered nature of FL development, from initial learning and prototyping to deployment, scaling, and operational sustainability.

\subsection{Evidence-Informed Recommendations}
Our findings identify recurring challenges reported in public Stack
Overflow discussions and GitHub artifacts. The following recommendations
synthesize these observed patterns and should be interpreted as
evidence-informed directions rather than validated best practices or
reference standards. Developer reports may reflect local environments,
project-specific constraints, incomplete diagnostic information, or
individual interpretations. Their applicability should therefore be
assessed within the context of each FL project.
\newline
\begin{itemize}
    \item \textbf{Environment Setup and Dependency Management: }We found that developers frequently struggle with installing and configuring FL toolchains due to dependency conflicts, version drift, and platform-specific constraints. Issues often arise when frameworks evolve rapidly, leading to breaking changes across APIs, incompatible library versions, and inconsistent behavior across local, cloud, and GPU environments. These challenges are amplified when users attempt to reproduce published results or follow outdated tutorials. Improvements such as versioned installation guides, backward-compatible APIs, standardized environment templates (e.g., Conda/Poetry/Docker), and automated compatibility checks can reduce setup friction and improve reproducibility.

    \item \textbf{Client Orchestration and Training Coordination: }We found that many FL implementations face coordination challenges in managing distributed clients, including client selection, straggler mitigation, partial participation, and synchronization across rounds. Questions and issues often reflect difficulties in running specific subsets of clients, handling unstable connectivity, and coordinating heterogeneous compute resources. Enhancements such as robust client scheduling policies, asynchronous or semi-synchronous training modes, adaptive round control, and better support for fault-tolerant execution can significantly improve training stability and practical deployability.

    \item \textbf{Data Partitioning, Non-IID Handling, and Evaluation: }A recurring difficulty concerns data heterogeneity and evaluation reliability. Developers frequently encounter unexpected behaviors when workloads are highly non-IID, when client datasets are imbalanced, or when label distributions differ substantially across clients. These conditions can trigger training instability, misleading metrics, or failures in evaluation pipelines (e.g., incorrect aggregation of metrics, missing test logic, or invalid assumptions about input dimensions). Improvements such as well-documented dataset partitioning utilities, standardized evaluation protocols, calibrated metrics reporting (global vs. per-client), and reproducible benchmark suites can reduce confusion and improve scientific validity.

    \item \textbf{Debugging, Error Messages, and Observability: }FL systems remain difficult to debug because errors may originate from multiple layers, including networking, serialization, cryptographic protocols, ML training loops, and distributed runtime orchestration. We observed that developers often face ambiguous error messages, silent failures, and difficulty identifying which client, round, or component triggered an error. To address this, FL frameworks should adopt structured logging, traceable experiment metadata, clear exception hierarchies, and built-in observability tools (e.g., per-round dashboards, client health reports, and reproducible failure traces). Improved documentation with common failure modes and troubleshooting checklists can further reduce debugging time.

    \item \textbf{Scalability, Performance, and Communication Efficiency: }We found that performance bottlenecks in FL commonly stem from communication overhead, inefficient aggregation, and resource imbalance across clients. Developers often report slow convergence, excessive round times, and limited scalability when simulating many clients or deploying cross-device FL. Techniques such as communication-efficient training (e.g., compression and partial updates), adaptive client sampling, aggregation optimizations, and hardware-aware scheduling can mitigate these bottlenecks. In addition, frameworks can provide practical guidance on scaling experiments (e.g., simulation best practices, batching strategies, and GPU utilization) to help practitioners avoid common performance pitfalls.

    \item \textbf{Privacy, Security, and Compliance-by-Design: }Since FL is frequently adopted to reduce direct data sharing, privacy and security assumptions must be explicit and enforceable. Developers often struggle to correctly integrate privacy-preserving mechanisms such as secure aggregation and differential privacy, and may be uncertain about threat models, privacy accounting, and deployment constraints across organizations. FL toolchains should provide secure-by-default configurations, clear documentation of supported threat models, validated privacy implementations, and testing utilities that verify privacy/security properties. Providing compliance-oriented guidance for cross-silo deployments (e.g., access control, auditability, and governance workflows) can also improve real-world adoption.
\end{itemize}

These recommendations identify potential directions for improving FL
tooling, documentation, reproducibility, and operational support.
However, their effectiveness has not been directly evaluated in this
study and requires further empirical validation.

\subsection{Implications for Software Engineering Practice and Research}

The recurring challenges identified from public Stack Overflow discussions
and popular open-source GitHub repositories provide evidence-informed
directions for improving the usability and maintainability of FL systems.
These implications are organized around the needs of practitioners,
framework developers and maintainers, educators, and researchers.

\paragraph{Implications for Practitioners.}
The frequent discussions of environment configuration, API compatibility,
data preparation, training, evaluation, and deployment show that many
practical FL difficulties arise from engineering and operational
conditions. Practitioners can mitigate these problems through versioned
dependencies, containerized environments, tested configurations, and
reproducible execution scripts. Standardized evaluation utilities,
actionable error messages, structured logs, and round-level traces can
also support more reliable diagnosis of distributed failures.

\paragraph{Implications for Framework Developers and Maintainers.}
The findings highlight opportunities for automated compatibility checks,
configuration validation, data-partition verification, and pre-execution
diagnostics. The prevalence of procedural discussions on Stack Overflow
suggests a need for task-oriented tutorials and executable examples,
whereas the stronger diagnostic intent observed on GitHub indicates a
need for clearer failure explanations, migration guidance, and
troubleshooting documentation. Structured issue templates requesting
framework versions, environments, commands, logs, and minimal
reproducible examples may further improve problem diagnosis and
maintenance efficiency.

\paragraph{Implications for Educators.}
FL education should complement algorithmic concepts with practical
software engineering skills, including dependency management, non-IID
data preparation, metric aggregation, distributed debugging, and privacy
configuration. Hands-on exercises involving heterogeneous clients,
partial participation, and realistic deployment failures can better
prepare students for the challenges encountered in operational FL
workflows.

\paragraph{Implications for Researchers.}
The identified topics and resolution outcomes provide a basis for
studying FL developer experience, framework evolution, and support
practices. Future studies can empirically evaluate whether improved
documentation, diagnostics, validation tools, and observability reduce
unresolved rates or resolution times. They can also extend the evidence
to smaller, proprietary, and domain-specific projects and develop
platform-specific taxonomies for bug reports, feature requests,
maintenance activities, and other repository communications.

\subsection{Broader Impacts of the Identified Developer Pain Points}
\label{sec:broader_impacts}

Although BERTopic provides the primary mechanism for organizing developer discussions, our interpretation is not based on topic modelling alone. We complement the identified topics with manual inspection, question-type classification, topic prevalence, unresolved or unaccepted-answer rates, median resolution time, temporal analysis, and cross-platform comparison. Together, these analyses reveal not only what FL developers discuss, but also the forms of support they seek and the challenges that remain difficult or time-consuming to resolve.

The findings indicate that recurring environment incompatibilities, dependency conflicts, API changes, configuration errors, and documentation gaps can create substantial maintenance and support burdens. Developers may depend on outdated releases, locally modified environments, or framework-specific workarounds, while maintainers repeatedly address similar compatibility and migration problems. These challenges can increase maintenance effort, fragment the framework ecosystem, and slow the adoption of newer and more reliable releases.

The identified data-partitioning, benchmark-reproduction, training-instability, model-state, and evaluation problems may also undermine scientific reproducibility and result validity. In FL, an experiment may execute successfully while using unintended client partitions, inconsistent preprocessing, incorrect model states, or improperly aggregated metrics. Such problems are especially consequential under non-IID data and partial client participation, where global and client-level performance can differ substantially. Similarly, orchestration, networking, container, GPU, runtime, and RPC failures can delay the transition from simulation to real multi-party deployment and make failures difficult to isolate across distributed components.

\begin{table*}[htbp]
\centering
\caption{Synthesis of identified FL developer challenges and their broader impacts.}
\label{tab:impact_synthesis}
\small
\setlength{\tabcolsep}{4pt}
\renewcommand{\arraystretch}{1.10}

\begin{tabularx}{\textwidth}{
>{\raggedright\arraybackslash}p{0.15\textwidth}
>{\raggedright\arraybackslash}p{0.25\textwidth}
>{\raggedright\arraybackslash}p{0.25\textwidth}
>{\raggedright\arraybackslash}X}
\toprule
\textbf{Impact area} &
\textbf{Supporting findings} &
\textbf{Potential impact} &
\textbf{Implications} \\
\midrule

Maintenance and framework evolution &
Environment incompatibilities, dependency conflicts, API breakages, configuration errors, and outdated documentation &
Repeated support effort, difficult migrations, fragmented environments, and continued reliance on obsolete releases &
Stable APIs, compatibility matrices, automated cross-version testing, executable examples, and migration guidance \\

Reproducibility and evaluation validity &
Data-partitioning problems, incomplete benchmark configurations, training instability, model-state errors, and incorrect metric aggregation &
Experiments may run successfully while producing misleading, incomparable, or irreproducible results &
Reference data partitions, explicit seeds and sampling policies, validated evaluation utilities, and reproducible benchmark configurations \\

Deployment and operational reliability &
Client orchestration, networking, Docker, Kubernetes, GPU, runtime, and RPC failures &
Delayed transition from simulation to deployment, increased operational effort, and difficult failure isolation &
Structured logging, client- and round-level tracing, health checks, fault-tolerant execution, and reproducible failure reports \\

Privacy and trustworthiness &
Difficulties integrating differential privacy, secure aggregation, privacy accounting, and privacy-preserving feature engineering &
Reduced model utility, incorrect threat-model assumptions, or weaker privacy protection than intended &
Validated privacy components, explicit threat models, automated privacy tests, and reference privacy configurations \\

Developer onboarding and adoption &
Procedural questions, installation barriers, documentation gaps, and uncertainty about framework scope &
Slower onboarding, repeated troubleshooting, and reduced adoption among teams without specialized FL expertise &
Task-oriented documentation, end-to-end examples, troubleshooting guides, and realistic educational exercises \\

\bottomrule
\end{tabularx}
\end{table*}

Difficulties involving differential privacy, secure aggregation, and privacy accounting further affect the trustworthiness of FL systems. Incorrect privacy configurations may substantially reduce model utility or provide weaker protection than intended. More broadly, the prevalence of procedural, setup, and troubleshooting questions suggests that developer experience directly influences framework adoption. The maturity of an FL framework should therefore be assessed not only by the algorithms it implements, but also by the stability of its interfaces, reproducibility of its experiments, observability of its execution, validity of its evaluation utilities, and usability of its privacy-preserving components.

Table~\ref{tab:impact_synthesis} summarizes how the recurring challenges identified across Stack Overflow and GitHub may affect the maintenance, reproducibility, deployment, trustworthiness, and adoption of FL systems.

These impacts are interconnected rather than isolated. For example, dependency and configuration problems increase maintenance effort while also undermining reproducibility and delaying deployment. Similarly, limited observability affects both operational reliability and the ability to validate experimental results. Addressing these challenges therefore requires coordinated improvements across framework design, documentation, testing, evaluation, privacy support, and runtime observability.

Because our study examines public Stack Overflow and GitHub artifacts, these impacts should be interpreted as analytically transferable patterns rather than causal or population-level estimates. Their prevalence and severity may differ across frameworks, organizations, application domains, and deployment settings. Nevertheless, the recurrence of related difficulties across both platforms provides complementary evidence that maintainability, reproducibility, deployment reliability, privacy assurance, and developer support are central to the continued evolution and adoption of FL systems.

\subsection{Interpreting FL Challenges through an Empirical Software Engineering Lens}
\label{sec:ese_interpretation}
The identified topics can be interpreted as manifestations of broader empirical software engineering challenges rather than as isolated FL-specific failures.

First, recurring installation problems, unclear configuration behaviour, incomplete documentation, and difficult-to-diagnose errors directly affect developer experience. Developer experience concerns how developers perceive and interact with their tools, processes, and working environments \cite{fagerholm2012developer}. In the FL context, developers must manage machine-learning libraries, distributed runtimes, communication protocols, privacy mechanisms, and heterogeneous hardware simultaneously. The repeated need for procedural guidance and troubleshooting therefore indicates substantial cognitive and operational effort before developers can focus on the intended learning task.

Second, the frequent occurrence of missing functions, renamed interfaces, incompatible dependencies, and outdated tutorials reflects the consequences of API evolution. Although API changes can support framework improvement, they also impose migration effort on downstream users and may invalidate examples, configurations, and integrations built for earlier versions \cite{robbes2012developers}. Our findings show that this burden is especially pronounced in FL because a change in one framework or dependency can propagate across model code, data pipelines, execution environments, and deployment configurations. API stability, deprecation policies, versioned documentation, and executable migration examples are therefore central maintenance concerns rather than only usability improvements.

Third, the findings expose stability challenges within the wider FL software ecosystem. FL applications depend on interconnected frameworks, numerical libraries, container images, communication middleware, orchestration platforms, privacy components, and hardware-specific toolchains. Software ecosystems combine technical dependencies with the communities that develop and maintain them \cite{mens2023introduction}. Consequently, failures such as dependency conflicts, unavailable packages, broken container builds, or inconsistent GPU configurations can propagate across project boundaries. The reliability of an FL framework therefore depends partly on the compatibility, maintenance activity, and documentation quality of its surrounding ecosystem.

Finally, the cross-platform findings illustrate the socio-technical nature of FL development. Stack Overflow primarily supports knowledge seeking and user-facing troubleshooting, whereas GitHub coordinates problem reporting, diagnosis, implementation, and maintenance. Resolving distributed failures often requires coordination among users, maintainers, infrastructure specialists, and contributors who possess different technical knowledge. From a socio-technical perspective, successful maintenance depends on alignment between technical dependencies and the communication structures used to manage them \cite{cataldo2008socio}. Long resolution times for deployment, runtime, and infrastructure topics may therefore reflect not only technical complexity but also the coordination required to reproduce problems, identify responsible components, and develop compatible fixes.

Overall, these connections show that FL developer pain points arise at the intersection of developer experience, software evolution, ecosystem dependencies, and collaborative coordination. Improving FL tooling consequently requires more than correcting individual defects: it requires stable interfaces, compatible dependency strategies, observable distributed execution, maintainable documentation, and support mechanisms that enable knowledge to move effectively between users and framework maintainers.
 
\section{Threats to Validity}\label{threats-to-validity}
Threats to validity refer to factors or influences that can compromise the accuracy, reliability, or generalizability of study findings. Validity of research is concerned with the question of how the conclusions might be wrong, i.e., the relationship between conclusions and reality \cite{bean2007qualitative}. These threats can introduce errors, biases, or limitations that affect the validity and credibility of research results. Identifying and addressing these threats is essential to ensure the robustness and trustworthiness of research findings. In our study, we acknowledge the presence of the following potential threats.

\textbf{External Validity and Sampling Bias:} refers to how well research results can be applied to other settings or broader groups beyond the studied data \cite{wohlin2012experimentation}. Our GitHub dataset consists of public, predominantly English-language
FL repositories with at least 300 stars. This threshold allowed us to
focus on established projects with visible development activity and
sufficient discussion data. However, it may underrepresent smaller,
newer, specialized, experimental, or less popular repositories. Such
projects may experience different challenges, particularly during early
development or within narrow application domains. GitHub stars also
measure project visibility and interest rather than software quality,
maintainer activity, or industrial adoption. The Stack Overflow corpus may omit relevant FL discussions that use only
generic tags such as \texttt{python}, \texttt{pytorch}, or
\texttt{tensorflow}. Because these tags contain large volumes of mostly
non-FL posts, we relied on FL-related tag co-occurrence and threshold
filtering; therefore, the findings apply to the captured public
discussions rather than all potentially relevant Stack Overflow posts.

The study is additionally limited to publicly available Stack Overflow
and GitHub artifacts. Developers working with proprietary FL systems or
within private organizational environments may encounter challenges
that are not publicly reported. Relevant discussions may also occur in
official framework forums, mailing lists, Slack or Discord communities,
private issue trackers, and direct communication with maintainers.
Restricting the analysis to English-language content may further exclude
experiences from other developer communities.

Although the dataset includes repositories from different FL frameworks
and application settings, the results are reported primarily at an
aggregate level. Challenges may differ across cross-device and
cross-silo FL, simulation and production deployment, and domains such
as healthcare, finance, mobile computing, and edge systems. The findings
should therefore be interpreted as patterns observed in prominent
public FL development communities rather than as estimates representing
all FL projects and practitioners.

\textbf{Internal Validity and Reliability:}

The identified topic structure may be influenced by preprocessing
choices, the selected sentence-embedding model, UMAP configuration,
HDBSCAN parameters, and the representation model used by BERTopic.
Alternative modelling decisions could therefore produce somewhat
different topic boundaries. To reduce this threat, we evaluated multiple
parameter configurations, selected the final models based on topic
coherence and interpretability, and manually inspected topic keywords
and representative artifacts. The topic labels were reviewed by
multiple authors with experience in FL and empirical software
engineering. We also provide the preprocessing, modelling, and analysis
scripts in the replication package to support reproducibility and
independent assessment.

Manual interpretation remains necessary because automatically generated
keywords may not fully represent the meaning of a cluster. We mitigated
this risk by examining representative posts, issues, and pull requests
rather than assigning labels from keywords alone. Nevertheless, some
artifacts may reasonably relate to more than one topic, whereas the
model assigns each artifact to a single dominant topic.

\textbf{Construct Validity:}Construct validity concerns whether the measures used in the study adequately represent the concepts being investigated \cite{wohlin2012experimentation}. We assessed topic
difficulty using the proportion of Stack Overflow posts without accepted
answers, the proportion of open GitHub issues and pull requests, and
median resolution time. These measures provide observable indications
of resolution difficulty and support availability, but they do not
directly measure the inherent technical complexity of a problem.

A Stack Overflow question may lack an accepted answer because the asker
did not return to accept an available solution, the question was unclear,
or the topic received limited community attention. Similarly, a GitHub
artifact may remain open because it has low priority, depends on external
work, lacks maintainer resources, or is no longer actively pursued.
Conversely, closure does not necessarily indicate that the reported
problem was successfully resolved. An issue may be closed as duplicated,
invalid, obsolete, or unsupported, while a pull request may be closed
without being merged. Resolution time may also reflect project
prioritization, release schedules, maintainer availability, and
coordination effort in addition to technical difficulty. We therefore
interpret unresolved rates and resolution times as complementary
indicators of observed resolution difficulty rather than definitive
measures of problem complexity.

The classification of question types introduces another construct
threat because some artifacts express multiple intentions or communicate
their main intention implicitly. To mitigate this concern, we manually
validated a stratified sample of 440 artifacts covering both platforms
and all identified topics. The consensus manual labels were compared
with the hybrid automatic classifications, producing a Cohen's kappa of
$\kappa=0.78$, which indicates substantial agreement.

Applying a common How--Why--What--Other taxonomy to Stack Overflow and
GitHub may simplify the broader communicative functions of repository
artifacts. In particular, GitHub issues and pull requests may combine
bug reporting, feature requests, maintenance, implementation, and
coordination. Although the \emph{Other} category partially accommodates
such artifacts, a broader platform-specific taxonomy could provide a
more detailed characterization. The RQ2 results should therefore be
interpreted as a comparison of dominant communicative intents rather
than a complete classification of all GitHub artifact purposes.

Additionally, we evaluated the difficulty of FL topics using the percentage of unresolved posts, issues, and pull requests, as well as the median time to resolution. While these measures are informative, they may not capture difficulty in all cases. Other indicators, such as median time to first answer, project priorities, maintainer resources, backlog management,
community participation, and user acceptance behaviour, number of answers per post (for Stack Overflow), and number of comments, issue labels, review comments, and review cycles (for GitHub), could also be explored. Nevertheless, the metrics we used are consistent with those applied in similar studies \citep{abdellatif2020challenges, ahmed2018concurrency, bajaj2014mining, rosen2016mobile, yang2016security}.

\textbf{Temporal Validity}

Resolution outcomes are influenced by the observation period. Artifacts
created earlier have had more time to receive an accepted answer or to
be closed than artifacts created near the end of data collection.
Unresolved artifacts are right-censored because their eventual outcome
and resolution time are unknown. Consequently, median resolution time
is calculated only from resolved artifacts and may not represent the
time that currently open artifacts will ultimately require.

Resolution patterns may also change as FL frameworks evolve. Major
releases, API deprecations, dependency updates, changes in documentation,
growth or decline in project activity, and variations in maintainer
availability can influence both unresolved rates and resolution times.
A long resolution time may therefore reflect the state of a framework
or community during a particular period rather than a persistent
property of the topic. Our temporal analysis provides context for topic
evolution, but the reported difficulty measures should be interpreted
within the study's data-collection window.

\textbf{Conclusion Validity}

Topic-level percentages and median resolution times may be less stable
for topics containing relatively few artifacts. We report topic sizes
alongside the corresponding measures and use medians to reduce the
influence of extreme resolution times. However, the results should be
interpreted together with the underlying sample sizes and the different
resolution mechanisms of Stack Overflow and GitHub. In particular, an
accepted Stack Overflow answer and a closed GitHub artifact represent
different forms of resolution and should not be treated as directly
equivalent outcomes. Accordingly, our cross-platform conclusions focus
on complementary patterns in developer support and maintenance rather
than assuming that the two platforms measure difficulty in exactly the
same way.

\section{Conclusion \& Future Work}\label{conclusion}
In this study, we carried out an empirical analysis to understand the challenges faced by FL developers. We mined and analyzed more than 495 Stack Overflow posts and 9,116 GitHub issues and pull requests to identify common problem areas, measure how difficult they are, and compare how developers ask questions across platforms. We found 9 main topics on Stack Overflow and 13 on GitHub. Many discussions focused on building and running FL workflows and managing distributed training. We also observed that most developer questions are ``How'' questions, showing that developers often need step-by-step guidance. Our difficulty analysis suggests that topics related to large-scale coordination, system integration, and API changes are among the hardest, often taking longer to resolve or remaining unresolved.

These results have important implications for FL framework developers, documentation writers, and researchers. They point to the need for better abstractions, stronger deployment support, and clearer documentation to reduce developer effort and confusion. Our approach also offers a simple and repeatable way to track developer pain points over time using public data, helping improve the usability, reliability, and maintainability of FL systems as they evolve. These findings should be interpreted as evidence of recurring challenges
within the analyzed public Stack Overflow and GitHub communities rather
than as representative estimates for all FL developers, proprietary
systems, or organizational settings.

Future research will explore the use of prompt-based large language
models (LLMs) to support FL development tasks such as client--server
pipeline setup, debugging, experiment configuration, and documentation
generation, with the goal of reducing the technical burden on framework
developers and practitioners. We also plan to broaden the evidence base
beyond Stack Overflow and GitHub by examining discussions from official
framework forums, Slack and Discord communities, mailing lists, and
benchmark or tutorial repositories.

In parallel, we will conduct interviews with FL researchers, framework
maintainers, and ML engineers to identify key pain points and practical
design requirements. Surveys of novice and experienced FL practitioners
will complement these qualitative findings by providing evidence on
onboarding barriers, evaluation and reproducibility difficulties, and
satisfaction with existing tooling. We also plan to conduct case studies
across representative FL application areas, including mobile and edge
learning, healthcare, and computer vision, to examine deployment
artifacts, configuration practices, and domain-specific challenges.
These studies will further enable empirical validation of the challenges
identified in this work and assess whether proposed interventions, such
as improved documentation, compatibility checking, configuration
validation, diagnostics, and observability, measurably reduce developer
effort and resolution time.

\section{Declarations}
\subsection*{Funding }This research is supported in part by the Natural Sciences and Engineering Research Council of Canada (NSERC), and by the industry-stream NSERC CREATE in Software Analytics Research (SOAR)

\subsection*{Ethical approval } Not Applicable

\subsection*{Informed consent }Not Applicable

\subsection*{Author Contributions}
\begin{itemize}
    \item \textbf{Sahand Saed} led the conceptualization and design of the study, performed the data collection and preprocessing, conducted the topic modeling and empirical analysis, and drafted the manuscript.
    \item \textbf{Khairul Alam }contributed to the study design, provided methodological guidance, reviewed and refined the data collection and topic analysis procedures, and critically revised the manuscript.
    \item \textbf{Banani Roy} supervised the overall research process, provided strategic direction for the study, contributed to the interpretation of results, refined topic analysis procedures, and helped with manuscript revision and finalization.
\end{itemize}

\subsection*{Data Availability Statements}
Our replication package can be found in our online appendix \citep{saed2026flchallenges}.

\subsection*{Conflict of Interest}
We have no conflict of interest.
\subsection*{Clinical Trial Number in the manuscript }Not Applicable
\subsection*{Acknowledgments}
This research is supported in part by the Natural Sciences and Engineering Research Council of Canada (NSERC), and by the industry-stream NSERC CREATE in Software Analytics Research (SOAR).
\bibliography{sn-bibliography}

@inproceedings{li2021understanding,
  title={Understanding quantum software engineering challenges an empirical study on stack exchange forums and github issues},
  author={Li, Heng and Khomh, Foutse and Openja, Moses and others},
  booktitle={2021 IEEE International Conference on Software Maintenance and Evolution (ICSME)},
  pages={343--354},
  year={2021},
  organization={IEEE}
}

@inproceedings{bajaj2014mining,
  title={Mining questions asked by web developers},
  author={Bajaj, Kartik and Pattabiraman, Karthik and Mesbah, Ali},
  booktitle={Proceedings of the 11th Working conference on mining software repositories},
  pages={112--121},
  year={2014}
}

@inproceedings{mcmahan2017communication,
  title     = {Communication-Efficient Learning of Deep Networks from Decentralized Data},
  author    = {McMahan, H. Brendan and Moore, Eider and Ramage, Daniel and Hampson, Seth and Ag{\"u}era y Arcas, Blaise},
  booktitle = {Proceedings of the 20th International Conference on Artificial Intelligence and Statistics (AISTATS)},
  year      = {2017}
}

@article{kairouz2021advances,
  title   = {Advances and Open Problems in Federated Learning},
  author  = {Kairouz, Peter and McMahan, H. Brendan and Avent, Brendan and Bellet, Aur{\'e}lien and Bennis, Mehdi and Bhagoji, Arjun Nitin and others},
  journal = {Foundations and Trends in Machine Learning},
  volume  = {14},
  number  = {1--2},
  pages   = {1--210},
  year    = {2021},
  doi     = {10.1561/2200000083}
}

@article{beutel2020flower,
  title   = {Flower: A Friendly Federated Learning Research Framework},
  author  = {Beutel, Daniel J. and Topal, Taner and Mathur, Akhil and Qiu, Xinchi and Fernandez-Marques, Javier and Gao, Yan and Sani, Lorenzo and Li, Kwing Hei and Parcollet, Titouan and Gusm{\~a}o, Pedro Porto Buarque de and Lane, Nicholas D.},
  journal = {arXiv preprint arXiv:2007.14390},
  year    = {2020}
}

@article{hoofnagle2019european,
  title={The European Union general data protection regulation: what it is and what it means},
  author={Hoofnagle, Chris Jay and Van Der Sloot, Bart and Borgesius, Frederik Zuiderveen},
  journal={Information \& communications technology law},
  volume={28},
  number={1},
  pages={65--98},
  year={2019},
  publisher={Taylor \& Francis}
}

@article{european2016regulation,
  title={Regulation (EU) 2016/679 (General data protection regulation)},
  author={European Parliament and Council},
  journal={Off J Eur Union},
  volume={119},
  pages={1--88},
  year={2016}
}

@article{litjens2017survey,
  title={A survey on deep learning in medical image analysis},
  author={Litjens, Geert and Kooi, Thijs and Bejnordi, Babak Ehteshami and Setio, Arnaud Arindra Adiyoso and Ciompi, Francesco and Ghafoorian, Mohsen and Van Der Laak, Jeroen Awm and Van Ginneken, Bram and S{\'a}nchez, Clara I},
  journal={Medical image analysis},
  volume={42},
  pages={60--88},
  year={2017},
  publisher={Elsevier}
}

@article{bonawitz2019towards,
  title={Towards federated learning at scale: System design},
  author={Bonawitz, Keith and Eichner, Hubert and Grieskamp, Wolfgang and Huba, Dzmitry and Ingerman, Alex and Ivanov, Vladimir and Kiddon, Chloe and Kone{\v{c}}n{\`y}, Jakub and Mazzocchi, Stefano and McMahan, Brendan and others},
  journal={Proceedings of machine learning and systems},
  volume={1},
  pages={374--388},
  year={2019}
}

@article{nguyen2021federated,
  title={Federated learning for internet of things: A comprehensive survey},
  author={Nguyen, Dinh C and Ding, Ming and Pathirana, Pubudu N and Seneviratne, Aruna and Li, Jun and Poor, H Vincent},
  journal={IEEE communications surveys \& tutorials},
  volume={23},
  number={3},
  pages={1622--1658},
  year={2021},
  publisher={IEEE}
}

@article{nguyen2022federated,
  title={Federated learning for smart healthcare: A survey},
  author={Nguyen, Dinh C and Pham, Quoc-Viet and Pathirana, Pubudu N and Ding, Ming and Seneviratne, Aruna and Lin, Zihuai and Dobre, Octavia and Hwang, Won-Joo},
  journal={ACM Computing Surveys (Csur)},
  volume={55},
  number={3},
  pages={1--37},
  year={2022},
  publisher={ACM New York, NY}
}

@misc{SentenceTransformers_allMiniLM_L6_v2,
  author       = {{Sentence-Transformers}},
  title        = {all-MiniLM-L6-v2},
  howpublished = {Hugging Face model card},
  year         = {2026},
  url          = {https://huggingface.co/sentence-transformers/all-MiniLM-L6-v2},
  urldate      = {2026-01-03}
}

@article{pedregosa2011scikit,
  title={Scikit-learn: Machine learning in Python},
  author={Pedregosa, Fabian and Varoquaux, Ga{\"e}l and Gramfort, Alexandre and Michel, Vincent and Thirion, Bertrand and Grisel, Olivier and Blondel, Mathieu and Prettenhofer, Peter and Weiss, Ron and Dubourg, Vincent and others},
  journal={the Journal of machine Learning research},
  volume={12},
  pages={2825--2830},
  year={2011},
  publisher={JMLR. org}
}

@misc{ScikitLearn_CountVectorizer_Docs,
  author       = {{scikit-learn developers}},
  title        = {sklearn.feature\_extraction.text.CountVectorizer documentation},
  year         = {2026},
  url          = {https://scikit-learn.org/stable/modules/generated/sklearn.feature\_extraction.text.CountVectorizer.html},
  urldate      = {2026-01-03}
}

@article{zhao2018federated,
  title={Federated learning with non-iid data},
  author={Zhao, Yue and Li, Meng and Lai, Liangzhen and Suda, Naveen and Civin, Damon and Chandra, Vikas},
  journal={arXiv preprint arXiv:1806.00582},
  year={2018}
}

@article{bean2007qualitative,
  title={Qualitative research design: An interactive approach},
  author={Bean, Cynthia J},
  journal={Organizational Research Methods},
  volume={10},
  number={2},
  pages={393},
  year={2007},
  publisher={SAGE PUBLICATIONS, INC.}
}

@article{alam2025empirical,
  title={An empirical investigation on the challenges in scientific workflow systems development},
  author={Alam, Khairul and Roy, Banani and Roy, Chanchal K and Mittal, Kartik},
  journal={Empirical Software Engineering},
  volume={30},
  number={5},
  pages={151},
  year={2025},
  publisher={Springer}
}

@article{alam2026analyzing,
  title={Analyzing GitHub Issues and Pull Requests in nf-core Pipelines: Insights into nf-core Pipeline Repositories},
  author={Alam, Khairul and Roy, Banani},
  journal={arXiv preprint arXiv:2601.09612},
  year={2026}
}

@article{alam2026maintenance,
  title={Maintenance and Support in Community-Driven Scientific Pipeline Ecosystems: A Cross-Platform Empirical Study of nf-core},
  author={Alam, Khairul and Roy, Kowsik and Rahman, Md Shamimur and Roy, Banani},
  journal={arXiv preprint arXiv:2607.10839},
  year={2026}
}

@article{jolliffe2016principal,
  title={Principal component analysis: a review and recent developments},
  author={Jolliffe, Ian T and Cadima, Jorge},
  journal={Philosophical transactions of the royal society A: Mathematical, Physical and Engineering Sciences},
  volume={374},
  number={2065},
  pages={20150202},
  year={2016},
  publisher={the Royal Society publishing}
}

@inproceedings{fagerholm2012developer,
  title={Developer experience: Concept and definition},
  author={Fagerholm, Fabian and M{\"u}nch, J{\"u}rgen},
  booktitle={2012 international conference on software and system process (ICSSP)},
  pages={73--77},
  year={2012},
  organization={IEEE}
}

@article{lalitha2019peer,
  title={Peer-to-peer federated learning on graphs},
  author={Lalitha, Anusha and Kilinc, Osman Cihan and Javidi, Tara and Koushanfar, Farinaz},
  journal={arXiv preprint arXiv:1901.11173},
  year={2019}
}

@inproceedings{robbes2012developers,
  title={How do developers react to API deprecation? The case of a Smalltalk ecosystem},
  author={Robbes, Romain and Lungu, Mircea and R{\"o}thlisberger, David},
  booktitle={Proceedings of the ACM SIGSOFT 20th International Symposium on the Foundations of Software Engineering},
  pages={1--11},
  year={2012}
}

@article{hegedHus2021decentralized,
  title={Decentralized learning works: An empirical comparison of gossip learning and federated learning},
  author={Heged{\H{u}}s, Istv{\'a}n and Danner, G{\'a}bor and Jelasity, M{\'a}rk},
  journal={Journal of Parallel and Distributed Computing},
  volume={148},
  pages={109--124},
  year={2021},
  publisher={Elsevier}
}

@article{Sheller2020_FLinMedicine,
  author  = {Sheller, Micah J. and Edwards, Brandon and Reina, G. Anthony and Martin, Jason and Pati, Sarthak and Kotrotsou, Aikaterini and Milchenko, Mikhail and Xu, Weilin and Marcus, Daniel and Colen, Rivka R. and Bakas, Spyridon},
  title   = {Federated learning in medicine: facilitating multi-institutional collaborations without sharing patient data},
  journal = {Scientific Reports},
  year    = {2020},
  volume  = {10},
  number  = {1},
  pages   = {12598},
  doi     = {10.1038/s41598-020-69250-1}
}

@article{kholod2020open,
  title={Open-source federated learning frameworks for IoT: A comparative review and analysis},
  author={Kholod, Ivan and Yanaki, Evgeny and Fomichev, Dmitry and Shalugin, Evgeniy and Novikova, Evgenia and Filippov, Evgeny and Nordlund, Mats},
  journal={Sensors},
  volume={21},
  number={1},
  pages={167},
  year={2020},
  publisher={MDPI}
}

@incollection{mens2023introduction,
  title={An introduction to software ecosystems},
  author={Mens, Tom and Roover, Coen De},
  booktitle={Software Ecosystems: Tooling and Analytics},
  pages={1--29},
  year={2023},
  publisher={Springer}
}

@inproceedings{Bonawitz2017_SecureAgg,
  author    = {Bonawitz, Keith and Ivanov, Vladimir and Kreuter, Ben and Marcedone, Antonio and McMahan, H. Brendan and Patel, Sarvar and Ramage, Daniel and Segal, Aaron and Seth, Karn},
  title     = {Practical Secure Aggregation for Privacy-Preserving Machine Learning},
  booktitle = {Proceedings of the 2017 ACM SIGSAC Conference on Computer and Communications Security (CCS)},
  year      = {2017},
  pages     = {1175--1191},
  doi       = {10.1145/3133956.3133982}
}

@inproceedings{Abadi2016_DPSGD,
  author    = {Abadi, Mart{\'\i}n and Chu, Andy and Goodfellow, Ian and McMahan, H. Brendan and Mironov, Ilya and Talwar, Kunal and Zhang, Li},
  title     = {Deep Learning with Differential Privacy},
  booktitle = {Proceedings of the 2016 ACM SIGSAC Conference on Computer and Communications Security (CCS)},
  year      = {2016},
  pages     = {308--318},
  doi       = {10.1145/2976749.2978318}
}

@inproceedings{cataldo2008socio,
  title={Socio-technical congruence: a framework for assessing the impact of technical and work dependencies on software development productivity},
  author={Cataldo, Marcelo and Herbsleb, James D and Carley, Kathleen M},
  booktitle={Proceedings of the Second ACM-IEEE international symposium on Empirical software engineering and measurement},
  pages={2--11},
  year={2008}
}

@article{Beutel2020_Flower,
  author  = {Beutel, Daniel J. and Topal, Taner and Mathur, Akhil and Qiu, Xinchi and Fernandez-Marques, Javier and Gao, Yan and Sani, Lorenzo and Li, Kwing Hei and Parcollet, Titouan and de Gusm{\~a}o, Pedro Porto Buarque and Lane, Nicholas D.},
  title   = {Flower: A Friendly Federated Learning Research Framework},
  journal = {arXiv preprint},
  year    = {2020},
  eprint  = {2007.14390},
  archivePrefix = {arXiv}
}

@article{He2020_FedML,
  author  = {He, Chaoyang and Li, Songze and So, Jinhyun and Zhang, Mi and Wang, Hongyi and Wang, Xiaoyang and Vepakomma, Praneeth and Singh, Abhishek and Qiu, Hang and Shen, Li and Zhao, Peilin and Kang, Yan and Liu, Yang and Raskar, Ramesh and Yang, Qiang and Annavaram, Murali and Avestimehr, Salman},
  title   = {FedML: A Research Library and Benchmark for Federated Machine Learning},
  journal = {arXiv preprint},
  year    = {2020},
  eprint  = {2007.13518},
  archivePrefix = {arXiv}
}

@article{Reina2021_OpenFL,
  author  = {Reina, G. Anthony and others},
  title   = {OpenFL: An open-source framework for Federated Learning},
  journal = {arXiv preprint},
  year    = {2021},
  eprint  = {2105.06413},
  archivePrefix = {arXiv}
}

@inproceedings{Zhang2018_TFbugs,
  author    = {Zhang, Yuhao and Chen, Yifan and Cheung, Shing-Chi and Xiong, Yingfei and Qin, Lu},
  title     = {An Empirical Study on TensorFlow Program Bugs},
  booktitle = {Proceedings of the 27th ACM SIGSOFT International Symposium on Software Testing and Analysis (ISSTA)},
  year      = {2018},
  pages     = {129--140},
  doi       = {10.1145/3213846.3213866}
}

@inproceedings{Humbatova2020_DLFaultTaxonomy,
  author    = {Humbatova, Nargiz and Jahangirova, Gunel and Bavota, Gabriele and Riccio, Vincenzo and Stocco, Andrea and Tonella, Paolo},
  title     = {Taxonomy of Real Faults in Deep Learning Systems},
  booktitle = {Proceedings of the ACM/IEEE 42nd International Conference on Software Engineering (ICSE)},
  year      = {2020},
  pages     = {1110--1121},
  doi       = {10.1145/3377811.3380395}
}

@inproceedings{Alshangiti2019_MLDevChallenges,
  author    = {Alshangiti, Moayad and Sapkota, Hitesh and Murukannaiah, Pradeep K. and Liu, Xumin and Yu, Qi},
  title     = {Why is Developing Machine Learning Applications Challenging? A Study on Stack Overflow Posts},
  booktitle = {2019 ACM/IEEE International Symposium on Empirical Software Engineering and Measurement (ESEM)},
  year      = {2019},
  pages     = {1--11},
  doi       = {10.1109/ESEM.2019.8870187}
}

@inproceedings{hamidi2021towards,
  title={Towards understanding developers’ machine-learning challenges: A multi-language study on stack overflow},
  author={Hamidi, Alaleh and Antoniol, Giuliano and Khomh, Foutse and Di Penta, Massimiliano and Hamidi, Mohammad},
  booktitle={2021 IEEE 21st International Working Conference on Source Code Analysis and Manipulation (SCAM)},
  pages={58--69},
  year={2021},
  organization={IEEE}
}

@article{liu2021fate,
  title   = {FATE: An Industrial Grade Platform for Collaborative Learning with Data Protection},
  author  = {Liu, Yang and Fan, Tao and Chen, Tianjian and Xu, Qian and Yang, Qiang},
  journal = {Journal of Machine Learning Research},
  volume  = {22},
  number  = {226},
  pages   = {1--6},
  year    = {2021}
}

@article{reina2021openfl,
  title   = {OpenFL: An Open-Source Framework for Federated Learning},
  author  = {Reina, Gustavo A. and Gruzdev, Alexander and Foley, Patrick and Perepelkina, Olga and others},
  journal = {arXiv preprint arXiv:2105.06413},
  year    = {2021},
  doi     = {10.48550/arXiv.2105.06413}
}

@article{rieke2020digitalhealth,
  title   = {The Future of Digital Health with Federated Learning},
  author  = {Rieke, Nicola and Hancox, Jason and Li, Wenqi and Milletari, Fausto and Roth, Holger R. and Albarqouni, Shadi and Bakas, Spyridon and others},
  journal = {npj Digital Medicine},
  volume  = {3},
  pages   = {119},
  year    = {2020},
  doi     = {10.1038/s41746-020-00323-1}
}

@book{vasiliev2020natural,
  title={Natural language processing with Python and spaCy: A practical introduction},
  author={Vasiliev, Yuli},
  year={2020},
  publisher={No Starch Press}
}

@book{hardeniya2016natural,
  title={Natural language processing: python and NLTK},
  author={Hardeniya, Nitin and Perkins, Jacob and Chopra, Deepti and Joshi, Nisheeth and Mathur, Iti},
  year={2016},
  publisher={Packt Publishing Ltd}
}

@article{mcinnes2018umap,
  title={Umap: Uniform manifold approximation and projection for dimension reduction},
  author={McInnes, Leland and Healy, John and Melville, James},
  journal={arXiv preprint arXiv:1802.03426},
  year={2018}
}

@article{li2020federated,
  title={Federated learning: Challenges, methods, and future directions},
  author={Li, Tian and Sahu, Anit Kumar and Talwalkar, Ameet and Smith, Virginia},
  journal={IEEE Signal Processing Magazine},
  volume={37},
  number={3},
  pages={50--60},
  year={2020},
  publisher={IEEE}
}

@article{he2020fedml,
  title={FedML: A research library and benchmark for federated machine learning},
  author={He, Chaoyang and Annavaram, Murali and Avestimehr, Salman and others},
  journal={arXiv preprint arXiv:2007.13518},
  year={2020}
}

@article{reimers2019sentence,
  title={Sentence-bert: Sentence embeddings using siamese bert-networks},
  author={Reimers, Nils and Gurevych, Iryna},
  journal={arXiv preprint arXiv:1908.10084},
  year={2019}
}

@misc{huggingface-models-2025,
  author       = {Hugging Face Community},
  title        = {Hugging Face Models},
  year         = {2025},
  howpublished = {\url{https://huggingface.co/models}},
  note         = {Accessed: 2025-10-07}
}

@article{mezinijumping,
  title={“Jumping Through Hoops”: Why do Java Developers Struggle With Cryptography APIs?},
  author={Mezini, Sarah Nadi1 Stefan Kr{\"u}ger2 Mira and Bodden, Eric},
  journal={GI-Edition},
  pages={57}
}

@inproceedings{ahmed2018concurrency,
  title={What do concurrency developers ask about? a large-scale study using stack overflow},
  author={Ahmed, Syed and Bagherzadeh, Mehdi},
  booktitle={Proceedings of the 12th ACM/IEEE international symposium on empirical software engineering and measurement},
  pages={1--10},
  year={2018}
}

@inproceedings{bagherzadeh2019going,
  title={Going big: a large-scale study on what big data developers ask},
  author={Bagherzadeh, Mehdi and Khatchadourian, Raffi},
  booktitle={Proceedings of the 2019 27th ACM joint meeting on european software engineering conference and symposium on the foundations of software engineering},
  pages={432--442},
  year={2019}
}

@inproceedings{abdellatif2020challenges,
  title={Challenges in chatbot development: A study of stack overflow posts},
  author={Abdellatif, Ahmad and Costa, Diego and Badran, Khaled and Abdalkareem, Rabe and Shihab, Emad},
  booktitle={Proceedings of the 17th international conference on mining software repositories},
  pages={174--185},
  year={2020}
}

@inproceedings{scoccia2021challenges,
  title={Challenges in developing desktop web apps: a study of stack overflow and github},
  author={Scoccia, Gian Luca and Migliarini, Patrizio and Autili, Marco},
  booktitle={2021 IEEE/ACM 18th International Conference on Mining Software Repositories (MSR)},
  pages={271--282},
  year={2021},
  organization={IEEE}
}

@inproceedings{dhasade2020towards,
  title={Towards prioritizing github issues},
  author={Dhasade, Akash Balasaheb and Venigalla, Akhila Sri Manasa and Chimalakonda, Sridhar},
  booktitle={Proceedings of the 13th Innovations in Software Engineering Conference (formerly known as India Software Engineering Conference)},
  pages={1--5},
  year={2020}
}

@article{jokhio2021mining,
  title={Mining GitHub Issues for Bugs, Feature Requests and Questions},
  author={Jokhio, Marvi},
  year={2021}
}

@incollection{campbell2015latent,
  title={Latent Dirichlet allocation: extracting topics from software engineering data},
  author={Campbell, Joshua Charles and Hindle, Abram and Stroulia, Eleni},
  booktitle={The art and science of analyzing software data},
  pages={139--159},
  year={2015},
  publisher={Elsevier}
}

@inproceedings{wang2019does,
  title={Where does LDA sit for GitHub?},
  author={Wang, Xukun and Lee, Matthias and Pinchbeck, Angie and Fard, Fatemeh},
  booktitle={2019 34th IEEE/ACM International Conference on Automated Software Engineering Workshop (ASEW)},
  pages={94--97},
  year={2019},
  organization={IEEE}
}

@article{grootendorst2022bertopic,
  title={BERTopic: Neural topic modeling with a class-based TF-IDF procedure},
  author={Grootendorst, Maarten},
  journal={arXiv preprint arXiv:2203.05794},
  year={2022}
}

@article{xu2021federated,
  title={Federated learning for healthcare informatics},
  author={Xu, Jie and Glicksberg, Benjamin S and Su, Chang and Walker, Peter and Bian, Jiang and Wang, Fei},
  journal={Journal of healthcare informatics research},
  volume={5},
  pages={1--19},
  year={2021},
  publisher={Springer}
}

@article{lim2020federated,
  title={Federated learning in mobile edge networks: A comprehensive survey},
  author={Lim, Wei Yang Bryan and Luong, Nguyen Cong and Hoang, Dinh Thai and Jiao, Yutao and Liang, Ying-Chang and Yang, Qiang and Niyato, Dusit and Miao, Chunyan},
  journal={IEEE communications surveys \& tutorials},
  volume={22},
  number={3},
  pages={2031--2063},
  year={2020},
  publisher={IEEE}
}

@article{guendouzi2023systematic,
  title={A systematic review of federated learning: Challenges, aggregation methods, and development tools},
  author={Guendouzi, Badra Souhila and Ouchani, Samir and Assaad, Hiba EL and Zaher, Madeleine EL},
  journal={Journal of Network and Computer Applications},
  pages={103714},
  year={2023},
  publisher={Elsevier}
}

@article{vucinich2023current,
  title={The current state and challenges of fairness in federated learning},
  author={Vucinich, Sean and Zhu, Qiang},
  journal={IEEE Access},
  year={2023},
  publisher={IEEE}
}

@article{shirvani2024federated,
  title={Federated Learning: Attacks, Defenses, Opportunities, and Challenges},
  author={Shirvani, Ghazaleh and Ghasemshirazi, Saeid and Beigzadeh, Behzad},
  journal={arXiv preprint arXiv:2403.06067},
  year={2024}
}

@article{jelodar2019latent,
  title={Latent Dirichlet allocation (LDA) and topic modeling: models, applications, a survey},
  author={Jelodar, Hamed and Wang, Yongli and Yuan, Chi and Feng, Xia and Jiang, Xiahui and Li, Yanchao and Zhao, Liang},
  journal={Multimedia tools and applications},
  volume={78},
  pages={15169--15211},
  year={2019},
  publisher={Springer}
}

@inproceedings{yi2009comparative,
  title={A comparative study of utilizing topic models for information retrieval},
  author={Yi, Xing and Allan, James},
  booktitle={Advances in Information Retrieval: 31th European Conference on IR Research, ECIR 2009, Toulouse, France, April 6-9, 2009. Proceedings 31},
  pages={29--41},
  year={2009},
  organization={Springer}
}

@book{wei2007topic,
  title={Topic models in information retrieval},
  author={Wei, Xing},
  year={2007},
  publisher={University of Massachusetts Amherst}
}

@article{yau2014clustering,
  title={Clustering scientific documents with topic modeling},
  author={Yau, Chyi-Kwei and Porter, Alan and Newman, Nils and Suominen, Arho},
  journal={Scientometrics},
  volume={100},
  pages={767--786},
  year={2014},
  publisher={Springer}
}

@article{xie2013integrating,
  title={Integrating document clustering and topic modeling},
  author={Xie, Pengtao and Xing, Eric P},
  journal={arXiv preprint arXiv:1309.6874},
  year={2013}
}

@inproceedings{luostarinen2013using,
  title={Using topic models in content-based news recommender systems},
  author={Luostarinen, Tapio and Kohonen, Oskar},
  booktitle={Proceedings of the 19th Nordic conference of computational linguistics (NODALIDA 2013)},
  pages={239--251},
  year={2013}
}

@article{choi2015improving,
  title={Improving performance of recommendation systems using topic modeling},
  author={Choi, Seongi and Hyun, Yoonjin and Kim, Namgyu},
  journal={Journal of Intelligence and Information systems},
  volume={21},
  number={3},
  pages={101--116},
  year={2015},
  publisher={Korea Intelligent Information System Society}
}

@inproceedings{bergamaschi2015comparing,
  title={Comparing LDA and LSA topic models for content-based movie recommendation systems},
  author={Bergamaschi, Sonia and Po, Laura},
  booktitle={Web Information Systems and Technologies: 10th International Conference, WEBIST 2014, Barcelona, Spain, April 3-5, 2014, Revised Selected Papers 10},
  pages={247--263},
  year={2015},
  organization={Springer}
}

@article{belwal2023extractive,
  title={Extractive text summarization using clustering-based topic modeling},
  author={Belwal, Ramesh Chandra and Rai, Sawan and Gupta, Atul},
  journal={Soft Computing},
  volume={27},
  number={7},
  pages={3965--3982},
  year={2023},
  publisher={Springer}
}

@inproceedings{eidelman2012topic,
  title={Topic models for dynamic translation model adaptation},
  author={Eidelman, Vladimir and Boyd-Graber, Jordan and Resnik, Philip},
  booktitle={Proceedings of the 50th Annual Meeting of the Association for Computational Linguistics (Volume 2: Short Papers)},
  pages={115--119},
  year={2012}
}

@inproceedings{asuncion2010software,
  title={Software traceability with topic modeling},
  author={Asuncion, Hazeline U and Asuncion, Arthur U and Taylor, Richard N},
  booktitle={Proceedings of the 32nd ACM/IEEE international conference on Software Engineering-Volume 1},
  pages={95--104},
  year={2010}
}

@inproceedings{zhai2011constrained,
  title={Constrained LDA for grouping product features in opinion mining},
  author={Zhai, Zhongwu and Liu, Bing and Xu, Hua and Jia, Peifa},
  booktitle={Advances in Knowledge Discovery and Data Mining: 15th Pacific-Asia Conference, PAKDD 2011, Shenzhen, China, May 24-27, 2011, Proceedings, Part I 15},
  pages={448--459},
  year={2011},
  organization={Springer}
}

@inproceedings{gethers2010using,
  title={Using relational topic models to capture coupling among classes in object-oriented software systems},
  author={Gethers, Malcom and Poshyvanyk, Denys},
  booktitle={2010 IEEE international conference on software maintenance},
  pages={1--10},
  year={2010},
  organization={IEEE}
}

@inproceedings{gokcimen2024topic,
  title={Topic Modelling Using BERTopic for Robust Spam Detection},
  author={Gokcimen, Tunahan and Das, Bihter},
  booktitle={2024 12th International Symposium on Digital Forensics and Security (ISDFS)},
  pages={1--5},
  year={2024},
  organization={IEEE}
}

@article{blei2003latent,
  title={Latent dirichlet allocation},
  author={Blei, David M and Ng, Andrew Y and Jordan, Michael I},
  journal={Journal of machine Learning research},
  volume={3},
  number={Jan},
  pages={993--1022},
  year={2003}
}

@article{blei2006correlated,
  title={Correlated topic models},
  author={Blei, David and Lafferty, John},
  journal={Advances in neural information processing systems},
  volume={18},
  pages={147},
  year={2006},
  publisher={MIT; 1998}
}

@inproceedings{blei2006dynamic,
  title={Dynamic topic models},
  author={Blei, David M and Lafferty, John D},
  booktitle={Proceedings of the 23rd international conference on Machine learning},
  pages={113--120},
  year={2006}
}

@article{lee2000algorithms,
  title={Algorithms for non-negative matrix factorization},
  author={Lee, Daniel and Seung, H Sebastian},
  journal={Advances in neural information processing systems},
  volume={13},
  year={2000}
}

@inproceedings{yan2013biterm,
  title={A biterm topic model for short texts},
  author={Yan, Xiaohui and Guo, Jiafeng and Lan, Yanyan and Cheng, Xueqi},
  booktitle={Proceedings of the 22nd international conference on World Wide Web},
  pages={1445--1456},
  year={2013}
}

@inproceedings{chen2023user,
  title={What do user experience professionals discuss online? Topic modeling of a user experience Q\&A community},
  author={Chen, Langtao},
  booktitle={International Conference on Human-Computer Interaction},
  pages={365--380},
  year={2023},
  organization={Springer}
}

@article{daud2010knowledge,
  title={Knowledge discovery through directed probabilistic topic models: a survey},
  author={Daud, Ali and Li, Juanzi and Zhou, Lizhu and Muhammad, Faqir},
  journal={Frontiers of computer science in China},
  volume={4},
  pages={280--301},
  year={2010},
  publisher={Springer}
}

@article{nie2017data,
  title={Data-driven answer selection in community QA systems},
  author={Nie, Liqiang and Wei, Xiaochi and Zhang, Dongxiang and Wang, Xiang and Gao, Zhipeng and Yang, Yi},
  journal={IEEE transactions on knowledge and data engineering},
  volume={29},
  number={6},
  pages={1186--1198},
  year={2017},
  publisher={IEEE}
}

@incollection{fiscus2002topic,
  title={Topic detection and tracking evaluation overview},
  author={Fiscus, Jonathan G and Doddington, George R},
  booktitle={Topic detection and tracking: event-based information organization},
  pages={17--31},
  year={2002},
  publisher={Springer}
}

@article{rosen2016mobile,
  title={What are mobile developers asking about? a large scale study using stack overflow},
  author={Rosen, Christoffer and Shihab, Emad},
  journal={Empirical Software Engineering},
  volume={21},
  pages={1192--1223},
  year={2016},
  publisher={Springer}
}

@inproceedings{alshangiti2019developing,
  title={Why is developing machine learning applications challenging? a study on stack overflow posts},
  author={Alshangiti, Moayad and Sapkota, Hitesh and Murukannaiah, Pradeep K and Liu, Xumin and Yu, Qi},
  booktitle={2019 ACM/IEEE International Symposium on Empirical Software Engineering and Measurement (ESEM)},
  pages={1--11},
  year={2019},
  organization={IEEE}
}

@article{yang2016security,
  title={What security questions do developers ask? a large-scale study of stack overflow posts},
  author={Yang, Xin-Li and Lo, David and Xia, Xin and Wan, Zhi-Yuan and Sun, Jian-Ling},
  journal={Journal of Computer Science and Technology},
  volume={31},
  pages={910--924},
  year={2016},
  publisher={Springer}
}

@inproceedings{mi2023identifying,
  title={Identifying Topics and Trends in DevOps: A Study of Stack Overflow Posts},
  author={Mi, Qing and Bao, Qinghang and Cui, Longjie},
  booktitle={2023 49th Euromicro Conference on Software Engineering and Advanced Applications (SEAA)},
  pages={402--409},
  year={2023},
  organization={IEEE}
}

@misc{saed2026flchallenges,
  author       = {Saed, Sahand and co-authors},
  title        = {FL\_Developers\_Challenges: Replication Package for
                  ``Developer Challenges on Federated Learning''},
  year         = {2026},
  howpublished = {\url{https://github.com/sahandsaed/FL\_Developers\_Challenges}},
  note         = {GitHub repository}
}

@article{zou2017towards,
  title={Towards comprehending the non-functional requirements through developers’ eyes: An exploration of stack overflow using topic analysis},
  author={Zou, Jie and Xu, Ling and Yang, Mengning and Zhang, Xiaohong and Yang, Dan},
  journal={Information and Software Technology},
  volume={84},
  pages={19--32},
  year={2017},
  publisher={Elsevier}
}

@article{barua2014developers,
  title={What are developers talking about? an analysis of topics and trends in stack overflow},
  author={Barua, Anton and Thomas, Stephen W and Hassan, Ahmed E},
  journal={Empirical software engineering},
  volume={19},
  pages={619--654},
  year={2014},
  publisher={Springer}
}

@inproceedings{bird2006nltk,
  title={NLTK: the natural language toolkit},
  author={Bird, Steven},
  booktitle={Proceedings of the COLING/ACL 2006 Interactive Presentation Sessions},
  pages={69--72},
  year={2006}
}

@article{mcinnes2017hdbscan,
  title={hdbscan: Hierarchical density based clustering.},
  author={McInnes, Leland and Healy, John and Astels, Steve and others},
  journal={J. Open Source Softw.},
  volume={2},
  number={11},
  pages={205},
  year={2017}
}

@book{wohlin2012experimentation,
  title={Experimentation in Software Engineering},
  author={Wohlin, Claes and Runeson, Per and H{\"o}st, Martin and Ohlsson, Magnus C. and Regnell, Bj{\"o}rn and Wessl{\'e}n, Anders},
  year={2012},
  publisher={Springer}
}

@inproceedings{treude2011programmers,
  title     = {How Do Programmers Ask and Answer Questions on the Web?},
  author    = {Treude, Christoph and Barzilay, Ohad and Storey, Margaret-Anne},
  booktitle = {Proceedings of the 33rd International Conference on Software Engineering (ICSE '11)},
  year      = {2011},
  pages     = {804--807},
  address   = {Waikiki, Honolulu, HI, USA},
  publisher = {Association for Computing Machinery (ACM)},
  doi       = {10.1145/1985793.1985907},
  isbn      = {978-1-4503-0445-0},
  note      = {NIER Track}
}

@article{han2020deeplearningframeworks,
  title        = {What do Programmers Discuss about Deep Learning Frameworks},
  author       = {Han, Junxiao and Shihab, Emad and Wan, Zhiyuan and Deng, Shuiguang and Xia, Xin},
  journal      = {Empirical Software Engineering},
  year         = {2020},
  volume       = {25},
  pages        = {2694--2747},
  publisher    = {Springer},
  doi          = {10.1007/s10664-020-09819-6}
}

@article{wan2019programmers,
  title={What do programmers discuss about blockchain? a case study on the use of balanced lda and the reference architecture of a domain to capture online discussions about blockchain platforms across stack exchange communities},
  author={Wan, Zhiyuan and Xia, Xin and Hassan, Ahmed E},
  journal={IEEE Transactions on Software Engineering},
  volume={47},
  number={7},
  pages={1331--1349},
  year={2019},
  publisher={IEEE}
}

\end{document}